\documentclass[showpacs,pra,aps,superscriptaddress,twocolumn]{revtex4}
\usepackage{mathrsfs}
\usepackage{color}
\usepackage{array}
\usepackage{amsmath}
\usepackage{graphicx}
\usepackage{amstext}
\usepackage{amsfonts}
\usepackage{amssymb}
\usepackage{bm}
\usepackage{epstopdf}

\newcommand{\Tr}[0]{\operatorname{Tr}}

\newcommand{\braket}[1]{\langle #1 \rangle}

\renewcommand{\t}[0]{\dagger}
\newcommand{\e}[0]{\mathrm{e}}
\newcommand{\w}[0]{\omega}

\newcommand{\brac}[1]{\left( #1 \right)}
\newcommand{\Brac}[1]{\left[ #1 \right]}

\begin{document}
\title{Non-Markovian decoherence dynamics of the hybrid quantum system \\ 
with a cavity strongly coupling to a spin ensemble: a master equation approach } 
\author{Kai-Ting Chaing}
\affiliation{Department of Physics and Center for Quantum Information Science,
National Cheng Kung University, Tainan 70101, Taiwan}
\author{Wei-Min Zhang}
\email{wzhang@mail.ncku.edu.tw}
\affiliation{Department of Physics and Center for Quantum Information Science,
National Cheng Kung University, Tainan 70101, Taiwan}

\begin{abstract}
Based on the recent experiments on the hybrid quantum system of a superconducting microwave cavity coupling 
strongly to an inhomogeneous broadening spin ensemble under an external driving field, we use the exact master equation approach 
to investigate its non-Markovian decoherence dynamics. Here the spin ensemble is made by negatively 
charged nitrogen-vacancy (NV) defects in diamond. Our exact master equation theory for open systems depicts 
 the experimental decoherence results and reveals the mechanism how the decoherence induced
 by the inhomogeneous broadening is suppressed in the strong-coupling regime. Moreover, we show how
 the spectral hole burning generates localized states to further suppress  
 the cavity decoherence. We also investigate the two-time correlations in this system to further show how 
quantum fluctuations manifest quantum memory.
\end{abstract}

\maketitle

\section{Introduction}
Cavity QED systems have become promising quantum devices for individual quantum system controlling
and for quantum information storage, processing and transmission \cite{Haroche2006}. 
Such systems are also attracted great interest  in the study of the fundamental quantum theory  
of open systems and the measurement-induced decoherence dynamics \cite{Carmichael93}.
Cavity QED is the study of interaction between matters and quantized electromagnetic fields. 
Since the early 1980s, the investigation of cavity QED has moved towards strong coupling systems with 
high quality factor cavities \cite{Kimble03}. The strong coupling cavity QED has been experimentally realized 
with a multitude of physical systems including Rydberg atoms in microwave cavities \cite{Haroche96,Walther01}, 
alkali atoms in optical cavities \cite{Kimble92,Kimble98,Kimble04}, and superconducting circuits \cite{Girvin04,Mooij04}, etc. 
For certain circumstance, in the strong-coupling regime, decoherence (coherence loss)
of photon and atom states can be effectively suppressed in cavity QED and single photon and single atom states can 
be dynamically manipulated.

Meanwhile, an alternative realization of strong-coupling cavity QED has also been developed with the hybrid quantum 
systems of  large  spin or atom ensembles coupled to superconducting microwave cavities \cite{Xiang13}, where the collective 
coupling strength is proportional to the square root of the number of emitters. Also the spin or atomic 
ensemble plays the role of quantum memories that quantum information can be coherently stored 
and retrieved at some later time. Such hybrid systems have been realized with various spin ensembles, 
such as negatively charged nitrogen-vacancy (NV) defects in diamond \cite{Majer11,Kubo11}, rare-earth spin 
ensembles \cite{Probst13}, and magnons in yttrium iron garnet \cite{Huebl13,Tabuchi14}, etc. In particular, 
the NV centers in diamond have the long coherence time even at room temperature, which shows the great potential for quantum information 
storage. The strong coupling of such spin ensembles with microwave cavities could make coherent transfer of 
quantum information more feasible.

However, due to the local magnetic dipole-dipole couplings of NV centers constituting the ensemble
to residual nitrogen paramagnetic impurities, the resonance line  of a large NV electron spin ensemble 
is inhomogeneously broadened \cite{Stanwix10,Kurucz11,Diniz11,Sandner11}.  Inhomogeneous spin ensemble broadening could induce
decoherence to the cavity system and therefore may limit the performance of the coherent transfer and 
storage of quantum information.  However, in the strong-coupling regime, inhomogeneous broadening induced 
decoherence may be suppressed upon the width and shape of the inhomogeneous broadening, which is 
called as "cavity protection effect" \cite{Kurucz11,Diniz11}.  Recently, Putz {\it et al.}~have experimentally demonstrated 
such effect through the real-time dynamics of a superconducting microwave cavity coupled strongly to the
spin ensemble of NV centers  \cite{Putz14,Putz17}. By appropriately choosing microwave pulses, they
can increase the amplitude of coherent oscillations between the cavity and the spin ensemble.
They also used a semi-classical mean field approximation to explain their observation  \cite{Krimer14,Krimer15}. 
In this paper, we will study 
the real-time dynamics of such hybrid quantum systems from the dynamics of open quantum systems with 
the exact master equation approach we developed \cite{Tu2008,Lei2012,Zhang2012,Zhang2018}. With this master 
equation formulation, some new insights of the decoherence dynamics in this hybrid quantum system can be revealed.

The experimental setup  \cite{Putz14, Putz17}  of the superconducting microwave cavity coupled strongly to the
spin ensemble of NV centers can be described by the Tavis-Cummings model \cite{TCmodel}. Therefore in this paper, 
we will start with the Tavis-Cummings model to investigate the decoherence dynamics of cavity QED due to the spin
ensemble. Such decoherence dynamics heavily involves non-Markovian memory processes. The Tavis-Cummings model
with the spin ensemble at arbitrary temperature cannot be solved exactly. However, the experimental setup enables one to
convert the spin ensemble into a bosonic ensemble using the Holstein-Primakoff approximation \cite{HP1938}, 
from which we can derive the exact master equation. Then we can systematically study the non-Markovian dynamics 
of Tavis-Cummings model using the master equation incorporating an external driving field 
\cite{Lei2012,Zhang2012}. We can also addresses rigorously quantum dissipation and thermal fluctuation together to depict
the decoherence dynamics. As a result, the non-Markovian decoherence dynamics in this hybrid quantum 
systems can be understood better. It reveals the mechanism of how the decoherence induced by the inhomogeneous broadening 
can be suppressed in the strong-coupling regime.
Also, we show that the solution of the decoherence suppression by the spectral hole burning is due to the  localized 
bound states (localized modes) between the cavity and the spin ensemble generated by the
hole burning spectral density. Localized states are dissipationless as a general property of open quantum systems \cite{Zhang2012}. 
We also use two time-correlation functions to explore quantum memory effect \cite{Ali2015,Ali2017,min-review}. 
It enables us to accurately address the dynamics  of the cavity photons in the quantum regime.

The rest of the paper is organized as follows. In Sec.~II, we discuss the master equation theory of open quantum systems
for the hybrid system studied in this paper. We will briefly discuss the generalized Tavis-Cummings model for the hybrid 
quantum system consisting of a superconducting microwave cavity coupled strongly to a spin ensemble of NV centers
under the control of microwave pulse driving field. With the Holstein-Primakoff approximation to the spin ensemble 
under the experimental setup, we obtain the exact master equation for the cavity system 
 with the external driving field. We also discuss the physical 
consequence of the dissipation and fluctuations from the master 
equation through the comparison with the exact quantum Langevin equation. We further explore how the two-time correlation functions 
characterize quantum memory. In Sec.~III and Sec.~IV, we study in detail the decoherence dynamics of the cavity field 
and compare with the experiment data and other theoretical methods for different setups of the spin ensemble
in this hybrid quantum system. We also use the time correlation functions to study quantum memory far from
the semiclassical dynamics in Sec. V. In the last, a conclusion is given in Sec. VI.

\section{Tavis-Cummings model with external driving field and the exact master equation method}
\label{TC}
The hybrid quantum system consists of a superconducting microwave cavity coupled strongly to the spin ensemble of NV centers
and the cavity is driven by an appropriate microwave pulse. This system can be described by a generalized Tavis-Cummings 
model with the following Hamiltonian,
\begin{align}
H(t) =  \,& \hbar \w_{c} a^{\t} a  +  \big[f(t) a^{\t}+ f^{*}(t) a \big] \notag \\
    & +  \sum_{k} \big[\Delta_{k}\sigma^{z}_{k} + V_{k} a^{\t} \sigma^{-}_{k} + V^{*}_{k} \sigma^{+}_{k} a\big] \notag \\
     & +   \sum_{l} \big[\hbar \w_{l} b^{\t}_{l} b_{l} + V_{l} a^{\t} b_{l} + V^{*}_{l} b^{\t}_{l} a \big].  \label{gtcm}
\end{align}
where the first term describes a single-mode cavity, and $a^{\t}$($a$) is the corresponding bosonic creation (annihilation) 
operator with $\w_{c}$ being the resonate frequency.   The second term describes the microwave pulse driving field
$f(t)=\eta(t)\mathrm{e}^{-i\w_p t} $ acting on the cavity with the time-dependent field strength $\eta(t)$ and the phase 
frequency $\w_{p}$. The driving pulse is used to control the dynamics of the  hybrid  system through the cavity.
The third term is the Hamiltonian of the spin ensemble of NV centers and its coupling with the cavity, where $\sigma^{z}_{k}, 
\sigma^{\pm}_k$ represent the three Pauli matrices for each spin with $\Delta_{k}$ being the corresponding two level energy 
splitting, and $g_{k}$ is the coupling amplitude.
The remaining Hamiltonian describes the cavity leakage into the free space electromagnetic background environment, where $c_{l}^{\t}$($c_{l}$) 
is the creation (annihilation) operator of the photonic mode $l$ with frequency $\omega_l$ in the background environment, and
$V_l$ is the coupling amplitude between the cavity mode and the background environmental mode $l$. 

The dynamics of the total systems (the cavity system, the spin ensemble and the free space environmental modes) is given 
by the total density matrix following the unitary evolution,
\begin{align}
\label{evolution}
\rho_{\text{tot}}(t)=U(t,t_0)\rho_{\text{tot}}(t_0)U^{\t}(t,t_0),
\end{align}
where $U(t,t_0)$=$\mathcal{T}\text{exp}\brac{\frac{1}{i\hbar}\int^{t}_{t_0}d\tau H(\tau)}$ is the time evolution operator 
of the total system, and ${\cal T}$ is the time-ordering operator. 
Initially, the cavity system is empty, the spin ensemble and the environment are in the partitioned thermal
state, that is, $\rho_{\text{tot}}(t_0)$=$\rho_{\text{c}}(t_0)$$\otimes$$\rho_{\text{s}}(t_0)$$\otimes$$\rho_{\text{e}}(t_0)$. Indeed, 
the density matrix of the cavity state $\rho_{\text{c}}(t_0)$ can be arbitrary, and the others are thermal equilibrium states:
$\rho_{\text{s}}(t_0)$=$\frac{1}{Z_s}e^{-\beta_{s} \sum_k\Delta_k \sigma^z_k}$ and $\rho_{\text{e}}(t_0)$=$\frac{1}{Z_e}
e^{-\beta_e \sum_l \hbar \omega_l b^\dag_l b_l}$, where $\beta_s=1/(kT_{s})$ and $\beta_e=1/(kT_{e})$ 
are the inverse of their initial temperatures which could be the same or different. Immediately after the time $t_0$, the  cavity, the spin ensemble and the 
environment evolve into a non-equilibrium state under Eq.~(\ref{evolution}). Because experimentally one measures the physical observables
of the cavity system, such as the cavity field intensity (photon numbers) and photon correlations, we shall focus on the cavity
photon state dynamics only, which is determined by the reduced density matrix 
\begin{align}
\rho_{c}(t)=\text{Tr}_{s+e}\big[\rho_{\text{tot}}(t)\big]
\end{align}
that encompasses all the influences caused by the spin ensemble and the environment after we trace over all all the spin and 
environmental states.

In the experimental setup  \cite{Putz14, Putz17}, only a small percentage of spins ($\approx\!\!10^{6}$) are excited 
in comparison with the total spin number ($\approx\!10^{12}$) in the ensemble, which corresponds to the high polarization 
spin ensemble \cite{Kurucz11}. Thus the Holstein-Primakoff approximation \cite{HP1938} 
\begin{align}
\sigma^z_k \equiv c^\t_k c_k -1/2, ~~ \sigma^+_k \equiv c^\t_k\sqrt{1-c^\t_k c_k} \simeq c^\t_k
\end{align}
can be applied to the spin variables, where $c^\t_k, c_k$ correspond to the creation and annihilation operators of 
the boson mode $k$.  As a result, the Hamiltonian Eq.~(\ref{gtcm}) can be reduced to  
\begin{align}
H =  \,& \hbar \w_{c} a^{\t} a  +  \big[f(t) a^{\t}+ f^{*}(t) a \big] \notag \\
    & +  \sum_{k} \big[\Delta_{k}c^\dag_{k} c_k + V_{k} a^\dag c_{k}  + V^{*}_{k} c^\dag_{k} a\big] \notag \\
     & +   \sum_{l} \big[\hbar \w_{l} b^{\t}_{l} b_{l} + V_{l} a^{\t} b_{l} + V^{*}_{l} b^{\t}_{l} a \big] ,   \label{gtcm}
\end{align}
with a good accuracy in the comparison with experimental data, as we will show later.
Then the dynamics of the hybrid quantum system (the cavity, the spin ensemble and the environment) can be exactly solved. 
Usually decoherence of photon states due to environmental leakages and atomic absorptions can be treated with 
constant decay rates (Markov decoherence phenomena) \cite{Carmichael93}.  For the strong coupling cavity QED system with the spin ensemble,
 the resonance line of the NV spin ensemble is inhomogeneously broadened so that non-Markovian decoherence
 dominates the dynamics in this hybrid quantum system. To study the non-Markovian decoherence dynamics of the system, 
 we apply the exact master equation approach developed recently \cite{Tu2008, Lei2012, Zhang2012, Zhang2018},
 which has been used in many other applications \cite{Lei2011,Yang2014,Lo2015,Ali2015,Yang2017,Yang2018,Lai2018}.

Following the standard procedure given in Ref.~\cite{Lei2012},  after traced over all the spin and 
environmental states from Eq.~(\ref{evolution}), we obtain the exact master  equation of the reduced density 
matrix for the cavity system governed by the total Hamiltonian of Eq.~(\ref{gtcm}),
\begin{align}
\label{M_eq}
\frac{d}{dt}\rho_{c}(t)  = & \frac{1}{i\hbar}\big[\w^{\prime}_{c}(t,t_0)a^{\t}a{+}f^{\prime*}(t,t_0)a{+}f^{\prime}(t,t_0)a^{\t},\rho_{c}(t) \big]\notag \\
                        & +\gamma(t,t_0) \big[2a\rho_{c}(t)a^{\t} \!-\! a^{\t}a\rho_{c}(t) \!-\! \rho_{c}(t)a^{\t}a \big] \notag \\
                        & +\widetilde{\gamma}(t,t_0) \big[a\rho_{c}(t)a^{\t} \! +\! a^{\t}\rho_{c}(t)a  \notag \\
		    &~~~~~~~~~~~~~~~~~~~~~~~~  \!-\! a^{\t}a\rho_{c}(t) \!-\!\rho_{c}(t)aa^{\t} \big],
\end{align}
where the first term describes a unitary evolution of cavity photon states, in which the effective cavity
frequency $\w^{\prime}_{c}(t,t_0)$ and the effective external driving field $f^{\prime*}(t,t_0)$ have taken into account
the renormalized effects nonpturbatively due to the coupling to 
the spin ensemble. The other two terms in the master equation represent the non-unitary evolution 
due to the dissipation and fluctuations that are induced completely by the the spin ensemble and the 
environment. The dissipation and fluctuations are characterized explicitly with the dissipation and 
fluctuation coefficients, $\gamma(t,t_0)$ and $\widetilde{\gamma}(t,t_0)$, respectively. They describe the detailed 
 relaxation of the cavity field and the thermalization between the cavity and the spin ensemble.
The most important feature in the above exact master equation is that the time nonlocal renormalized cavity
frequency $\w^{\prime}_{c}(t,t_0)$, renormalized external driving field $f^{\prime*}(t,t_0)$, 
dissipation coefficient $\gamma(t,t_0)$  and fluctuation coefficient $\widetilde{\gamma}(t,t_0)$ are determined 
nonperturbatively by the nonequilibrium Green functions 
\cite{Lei2012},
\begin{subequations}
\label{coeff}
\begin{align}
&i \w_{c}^{\prime}(t,t_0) + \gamma(t,t_0) = - \frac{\dot{u}(t,t_0)}{u(t,t_0)} ,  \label{dissic} \\
&f^{\prime}(t,t_0)=i\dot{y}(t,t_0)-i\Brac{\frac{\dot{u}(t,t_0)}{u(t,t_0)}y(t,t_0)}, \\
&\tilde{\gamma}(t,t_0)=\lim_{\tau \rightarrow t}\frac{\partial {v}(\tau,t)}{\partial \tau}-\Brac{\frac{\dot{u}(t,t_0)}{u(t,t_0)}v(t,t)+\text{c.c}} ,
\end{align}
\end{subequations}
where the nonequilibrium Green functions $u(t,t_0)$, $y(t,t_0)$ and $v(\tau,t)$ obey the following integro-differential equations
\begin{subequations}
\label{IDE}
\begin{align}
& \frac{d}{dt} u(t,t_0) \! + \!  i\w_{c} u(t,t_0)
    +  \!\! \int^{t}_{t_0}\!\!\!d\tau g(t,\tau)u(\tau,t_0)=0,\\
& \frac{d}{dt} y(t,t_0)\!+\! i\w_{c} y(t,t_0)  
    +   \!\!  \int^{t}_{t_0}  \!\! \! d\tau g(t,\tau)y(\tau,t_0)
    =\frac{1}{i\hbar}f(t) ,   \label{fte} \\
& \frac{d}{d\tau} v(\tau,t) \!+\! i\w_{c} v(\tau,t)
    +   \!\!  \int^{\tau}_{t_0}  \!\! \! d\tau^{\prime} g(\tau,\tau^{\prime})v(\tau^{\prime},t) \notag \\
    & ~~~~~~~~~~~~~~~~~~~~~~~~
       =  \!\!  \int^{t}_{t_0}  \!\! \! dt^{\prime} \widetilde{g}(\tau,t^{\prime})\bar{u}(t^{\prime},t), 
       ~~(\tau\leq t)  \label{vte}  
\end{align}
\end{subequations}
subjected to the boundary conditions $u(t_0,t_0)$=$1$,  $y(t_0,t_0)$=$0$ and $v(t_0,t)$=$0$. 
In Eq.~(\ref{IDE}), the nonlocal integral kernels, $g(t,\tau)$ and $\widetilde{g}(t,\tau)$, are given as follows,
\begin{subequations}
\begin{align}
\label{g}
g(t,\tau)&= \frac{1}{\hbar^2} \! \sum_{k} |V_{k}|^{2}\e^{-i\Delta_{k}(t-\tau)/\hbar}  + \frac{1}{\hbar^2} \! \sum_{l} |V_{l}|^{2}\e^{-i\w_{l}(t-\tau)}\notag \\
            & = \sum_{\alpha=s,e} \int_{0}^{\infty} \frac{d\w}{2\pi} J_{\alpha}(\w)\e^{-i\w \brac{t-\tau}}, \\
\label{g_tilde}
\widetilde{g}(t,\tau)&=  \frac{1}{\hbar^2} \! \sum_{k} |V_{k}|^{2} \bar{n}(\Delta_k/\hbar,T_s) \e^{-i\Delta_{k}(t-\tau)/\hbar}  \notag \\
& ~~~ + \frac{1}{\hbar^2} \! \sum_{l} |V_{l}|^{2} \bar{n}(\w_l,T_e) \e^{-i\w_{l}(t-\tau)} \notag \\
            & = \sum_{\alpha=s,e} \int_{0}^{\infty} \frac{d\w}{2\pi} J_{\alpha}(\w)\bar{n}(\w,T_{\alpha})\e^{-i\w \brac{t-\tau}},
\end{align}
\end{subequations}
where 
\begin{subequations}
\begin{align}
J_{s}(\w) = \frac{2\pi}{\hbar^2} \sum_k |V_k|^2 \delta(\omega- \Delta_k/\hbar) , \label{sesd}\\
J_{e}(\w) = \frac{2\pi}{\hbar^2} \sum_l |V_l|^2 \delta(\omega- \omega_l) \simeq 2\kappa 
\end{align}
\end{subequations}
are the spectral densities associated with the spin ensemble or the environment, respectively,
and $\bar{n}_{s,e}(\w,T_{s,e}) = 1/(e^{\hbar \omega/kT_{s,e}} -1)$ are the corresponding particle 
distribution function. The frequency-dependence of $J_s(\omega)$ depicts the inhomogeneous 
broadening of the spin ensemble spectrum, and the cavity decay rate $\kappa$ characterizes the 
cavity leakage effect.

From the above result, we can further simplify the integro-differential equation of  $u(t,t_0)$ as
\begin{align}
 \frac{d}{dt} u(t,t_0)  +&  ( i\w_{c} + \kappa) u(t,t_0)  = \notag  \\
    - & \!\! \int^{t}_{t_0}\!\!\!d\tau \!\!  \int_{0}^{\infty} \!\! \frac{d\w}{2\pi} J_{s}(\w)\e^{-i\w \brac{t-\tau}})u(\tau,t_0) . \label{rgf}
\end{align}
The result shows that without the spin ensemble, the solution of $u(t,t_0)$ simply describes the 
spontaneous decay of the cavity field.  The last term in Eq.~(\ref{rgf}) determines the
dynamical process of the cavity photons dissipating into the spin ensemble, as will be seen clearer later. 
Furthermore, because of the boundary conditions $y(t_0,t_0)$=$0$ and $v(t_0,t)$=$0$,  
Eqs.~(\ref{vte})-(\ref{fte}) can be analytically solved in terms of $u(t,t_0)$ \cite{Lei2012},
\begin{subequations}
\label{yvt}
\begin{align}
&y(t,t_0)= \frac{1}{i\hbar} \int^{t}_{t_0} \!\! d\tau u(t,\tau) f(\tau) ,  \label{edf} \\
&v(\tau, t) = \int^{\tau}_{t_0} \!\!\! dt_1 \!\! \int^{t}_{t_0} \!\!\! dt_2 u(\tau,t_1) \widetilde{g}(t_1,t_2) u^\dag(t,t_2) .  \label{vtt}
\end{align}
\end{subequations}
As we will see later from Eq.~(\ref{cavity-f}), $y(t,t_0)$ is the driving-field-induced cavity field combining with the dissipation effect 
raised from the spin ensemble and the environment. More precisely speaking,  $y(t,t_0)$ comes from the driving field 
directly and then decays mainly due to the coupling of the cavity with the spin ensemble.  While,  the two-time function
$v(\tau,t)$ is the correlation associated with the initial state of the spin ensemble and also the free-space environment. 
If the spin ensemble and the environment 
can be  initially prepared at zero temperature, then $v(\tau,t)=0$. Therefore, $v(\tau,t)$ is the quantum correlation of the cavity photon
mixed with thermal fluctuations. 
 
 Now, using the exact master equation, we can easily compute the
 experimentally measured cavity field $\langle a(t) \rangle = \Tr[a \rho_c(t)]$ or more precisely the cavity intensity 
 (cavity photon number) $\langle n(t) \rangle = \Tr[a^\dag a \rho_c(t)]$:
\begin{subequations}
 \begin{align}
 \langle a(t) \rangle = & u(t,t_0) \langle a(t_0) \rangle + y(t,t_0),  \\
 \langle n(t) \rangle  =& |u(t,t_0)|^2 \braket{a^{\t}(t_0)a(t_0)} +|y(t,t_0)|^2 +v(t,t)  \notag  \\
  &  + [u^{*}(t,t_0)y(t,t_0)\langle a^{\t}(t_0)\rangle + {\rm c.c.} ] \label{cavityi}
 \end{align}
\end{subequations}
The semiclassical method Putz {\it et al.} used based on the Volterra integral equation
to explain their experimental observations \cite{Putz14, Putz17} can be equivalently written as 
\begin{align}
 \langle n(t) \rangle_{\rm sc} = | \langle a(t) \rangle|^2 =  |u(t,t_0)\langle a(t_0)\rangle|^2  +|y(t,t_0)|^2 \notag  \\
    + [u^{*}(t,t_0)y(t,t_0)\langle a^{\t}(t_0)\rangle + {\rm c.c.} ] .
\end{align}
The difference between the full quantum mechanical solution given here and the semiclassical solution in Refs.~\cite{Putz14, Putz17} is
the quantum fluctuation 
\begin{align}
 \langle n(t) \rangle - & \langle n(t) \rangle_{\rm sc} =  v(t,t) \notag \\
   + &   |u(t,t_0)|^2( \langle a^{\t}(t_0)a(t_0)\rangle 
 -  \langle a^\dag(t_0)\rangle  \langle a(t_0)\rangle )  .
 \end{align}
 which is negligible in the semiclassical regime (the average photon number
$ \langle n(t) \rangle \sim 10^6$  as was measured in experiments  \cite{Putz14, Putz17}). 
However, in the quantum regime where $\langle n(t) \rangle $ is the order of one or less, 
quantum fluctuations becomes significant and the semiclassical method used  in  \cite{Putz14, Putz17}
becomes invalid. This effect should become particularly important in the study of quantum memory for quantum 
information processing, as we will discuss in Sec.~V.

For a self-consistency check and also for a clearer physical interpretation to the decoherence dynamics described 
in the master equation, the above result can also be obtained from the quantum Langevin equation
derived from Heisenberg equation of motion.  Using the Heisenberg equation of motion and eliminating
the variables of the spin ensemble as well as the free-space environmental modes, we have derived the following
quantum Langevin equation \cite{Yang2014,Tan2011,Yang2015,Yang2017} for the cavity field operator $a(t)$
\begin{align}
\label{A2}
\frac{d}{dt} & a(t)  + (i\w_{c} + \kappa) a(t)   \notag   \\
           &  + \!\! \int^{t}_{t_0} \!\! d\tau \!\!  \int_{0}^{\infty} \!\!\! \frac{d\w}{2\pi} J_{s}(\w)\e^{-i\w \brac{t-\tau}} a(\tau)       
                  = \frac{1}{i\hbar} \big[ f(t) +  \xi(t) \big], 
\end{align}
where $\xi(t)$ is the noise force induced by the spin ensemble and also the environment:
\begin{align}
\xi(t)=  \! \sum_{k} & V_{k}\e^{-i\Delta_{k}(t-t_0)/\hbar}c_{k}(t_0) 
              + \! \sum_{l}  V_{l}\e^{-i\w_{l}(t-t_0)}b_{l}(t_0) .
\end{align}

Moreover, due to the linearity of Eq.~(\ref{A2}), the general solution of the quantum Langevin equation is 
\begin{align}
a(t)=u(t,t_0)a(t_0)+y(t,t_0)+ \chi(t,t_0) ,  \label{cavity-f}
\end{align}
from which we can find that $u(t,t_0)$ and $y(t,t_0)$ are given by Eq.~(\ref{rgf})-(\ref{edf}), and $\chi(t,t_0)$ is the noise field induced by the noise force $\xi(t)$:
\begin{align}
\chi(t,t_0)= \frac{1}{i\hbar} \int^{t}_{t_0}d\tau u(t,\tau) \xi(\tau) 
\end{align}
It is easy to check that the correlation function $v(t,t)$ of Eq.~(\ref{vtt}) obtained in the master equation method is indeed the noise correlation:
\begin{align}
v(\tau,t) = \langle \chi^\dag (t,t_0) \chi(\tau,t_0) \rangle ,
\end{align}
at $\tau=t$. Thus, we recover the results from the master equation. 
These results show that all the time nonlocal parameters in the master equation are fully determined by 
the cavity's Green function $u(t,t_0)$ from Eq.~(\ref{rgf}). 
It should be pointed out that it is the convolution integral in Eq.~(\ref{rgf}) that takes into account all the possible non-Markovian 
 effects of the cavity system interacting with the spin ensemble \cite{Zhang2018}.  

\section{Decoherence dynamics of the cavity coupled strongly to the spin ensemble}
\label{section2.3}

In the two recent experimental setup by Putz {\it et al.}  \cite{Putz14, Putz17},  the  superconducting microwave cavity frequency $\w_c$
is set to $2\pi$$\times$$2.69$ GHz,  and the main frequency $\w_s$
of the spin spectral density is resonant with the cavity, $\w_c=\w_s$, but has the broadening effect.  To reduce the thermal fluctuation effect, 
the entire experimental setup is cooled down to $T_s=25$ mK, which is only about one-fifth of the excitation energy of spins. Besides, 
spins are surrounded by a Helmholtz coil which supplies a strong magnetic field to modify the spins in a relative ground state. 
By controlling the strength of the driving pulse, the number of injected photons is manifested to be about or less than $10^{6}$, 
which is much lower than the total number of spins $\approx$ $10^{12}$ in the cavity so that the Holstein-Primakoff approximation 
to the spin ensemble becomes applicable.  By applying the driving field, the cavity photonic state may be coherently stored 
and retrieved from the spin ensemble for quantum information processing. 

\begin{figure}[h]
\includegraphics[scale=0.55]{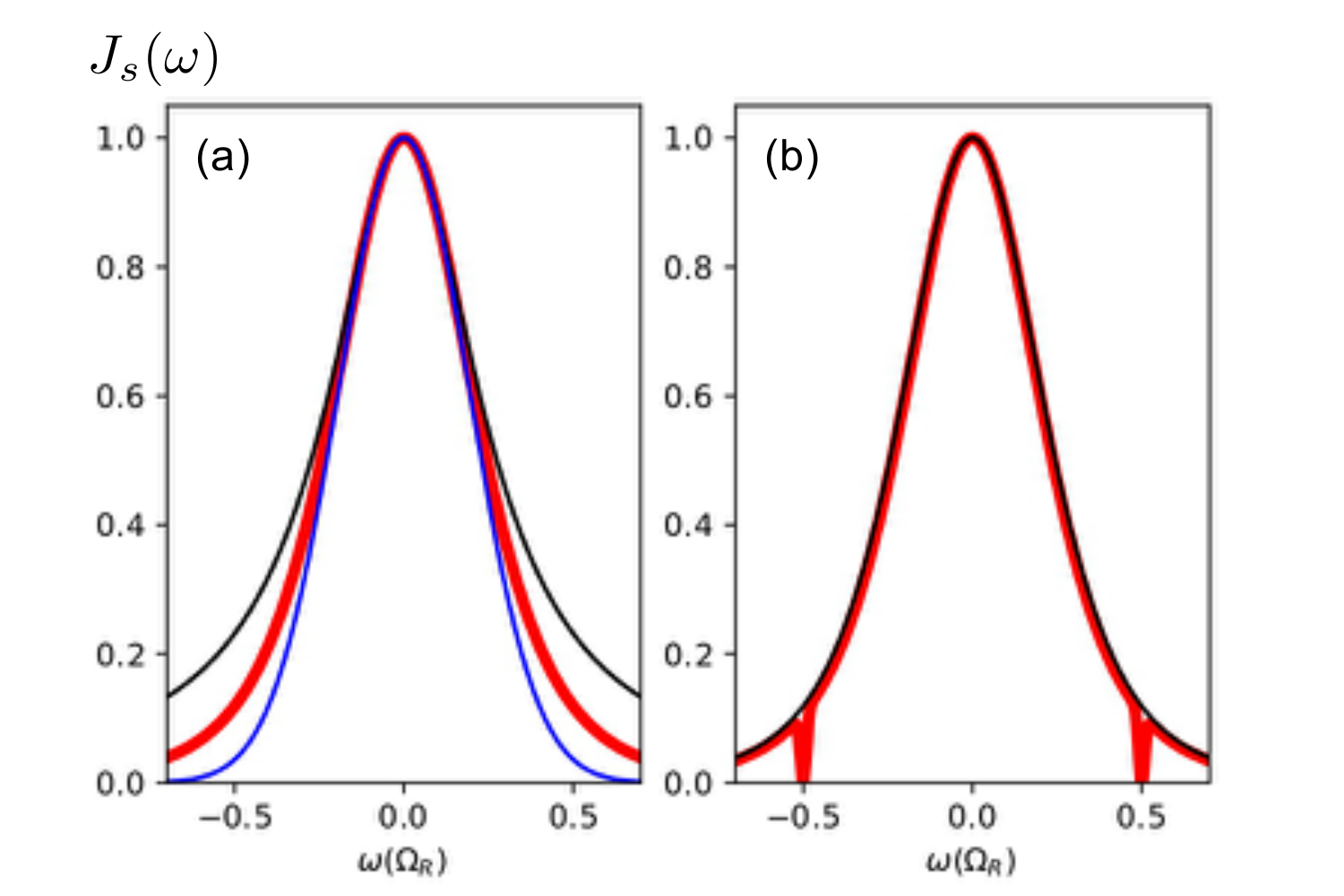}
\caption{The spectral density of the spin ensemble as a function of the frequency in the unit of $\Omega_R$ by a central frequency shifting $\w-\w_s \rightarrow \w$.
(a) The black, blue and red curves correspond to the spectral densities of a Gaussian form, a Lorentzian form, and the q-Gaussian 
form of Eq.~(\ref{sesd1}), respectively. (b) The black (thin) curve is the original q-Gaussian form, and the red (thick) curve is modified from the q-Gaussian 
form by the hole burning technique \cite{Putz17}.}
\label{J_qG}
\end{figure}

 The main difference between \cite{Putz14} and \cite{Putz17} is the manipulation of the broadening spectral density of the spin ensemble. The spectral 
 density of Eq.~(\ref{sesd})  is experimentally fitted as a q-Gaussian spectral density \cite{Putz14}
\begin{align}
J_s(\w)=2\pi \Omega^{2} \cdotp C \cdotp            
                    \Brac{1-(1-q)\frac{(\w-\w_{s})^2}{\Delta^2}}^{\frac{1}{1-q}},  \label{sesd1}
\end{align}
which is an intermediate form between a Gaussian spectral density and a Lorentzian spectral density with $q=1.39$, where C is its normalization 
constant, and $\w_{s}$ is the main frequency of the spin ensemble, $\Delta$ is determined by the full-width at the half maximum 
of  $J_s(\w)$ which is given by $\gamma$=$2\Delta\sqrt{\frac{2^{q}-2}{2q-2}}$=$18.8\pi$ MHz, and the coupling strength $2 \Omega$=$2\pi \times 17.2$ MHz represents 
a strong coupling in Ref.~\cite{Putz14}. The specific profile of the spectral density $J_s(\w)$ is shown in Fig.~\ref{J_qG}(a).
While the spectral density in the experiment \cite{Putz17} is a modification of Eq.~(\ref{sesd1}) as shown in Fig.~\ref{J_qG}(b), which is made by the 
spectral hole burning technique, a well-established technique in quantum optics used to turn off the excitations at some specific
frequencies in the spin ensemble. Here, the particular frequencies of $\w_s$$\pm$$\frac{\Omega_R}{2}$ are burnt out as shown 
in Fig.~\ref{J_qG}(b). Besides,  in 
the experiment \cite{Putz17}, the full-width at half maximum of the spectral density is changed slightly to  $18.2\pi$ MHz, and 
the coupling strength $\Omega\approx\frac{\Omega_R}{2}{=}21.3\pi$ MHz with $\Omega_R$ being the Rabi frequency. 
The cavity decay rate $\kappa/2\pi = 0.4$ MHz. Our calculations are all based on these experimental parameter setup.

The full quantum mechanical description of the hybrid quantum system given in the last section shows that the cavity dissipation or
decoherence dynamics is fully determined by the Green function $u(t,t_0)$ of Eq.~(\ref{rgf}).  Its general solution has been given
in \cite{Zhang2012}. Explicitly, taking a modified Laplace transformation $U(z)=\int^{\infty}_{t_0} u(t,t_0) \e^{iz (t-t_0)} dt$ to
Eq.~(\ref{rgf}), we have
\begin{equation}
U(z)=\frac{i}{z-\w_c + i\kappa -\Sigma(z)},   \label{lput}
\end{equation}
where $\Sigma(z)$ is the self-energy correction,
\begin{equation}
\label{Sigma_z}
\Sigma(z)=  \int^{\infty}_{-\infty} \frac{d\w}{2\pi} \frac{J_{s}(\w)}{z-\w},
\end{equation}
and
\begin{equation}
\label{Sigma_w}
\lim_{z\to\w\pm i\epsilon}\Sigma(z) =\Delta(\w)\mp \frac{i}{2}\! J_{s}(\w).
\end{equation}
Here $\Delta(\w)$=$\sum_{\alpha}\mathcal{P}\int^{\infty}_{-\infty} \frac{d\w^{\prime}}{2\pi} \frac{J_{\alpha}(\w^{\prime})}{\w-\w{\prime}}$ 
is the principal value of Eq.(\ref{Sigma_z}), which gives the cavity frequency shift due to the spin ensemble. 
The general solution to $u(t,t_0)$ is expressed as \cite{Zhang2012}
\begin{align}
\label{u_ana}
u(t, t_0)=& \sum_{\w_{b}} \mathcal{Z}(\w_{b})\e^{-i\w_{b}(t-t_0)}  \notag \\
&    +\frac{2}{\pi} \! \int_{-\infty}^{\infty}\! \! d\w\frac{J_s(\w)\e^{-i\w (t-t_0)}}
    {4\big[\w-\w_c-\Delta(\w)+i\kappa\big]^{2} +J^2_s(\w)}.
\end{align}
where the first term is a dissipationless term contributed from localized modes and the  
residues $\mathcal{Z}_{\w_{b}}{=}\frac{1}{1-\Sigma^{\prime}(\w_{s})}$ are the corresponding wave function amplitudes. 
A localized mode can exist only when the spectrum density has gaps or is a finite band, that is, $\sum_{\alpha} J_{\alpha}(\w_b)$=$0$ 
such that $\w_{b}{=}\w_{c}{+}\Delta(\w_{b})$. The second term of Eq.~(\ref{u_ana}) is a contribution from the branch cut due to the 
discontinuity of $\Sigma(z)$, so does $U(z)$, across the real axis on the complex space $z$ in the inverse Laplace transformation 
of Eq.~(\ref{lput}). It is shown by Eq.~(\ref{Sigma_w}). 
This branch cut  induces an non-exponential decay. Both contributions in Eq.~(\ref{u_ana}) are evidences of non-Markovian dynamics
in open quantum systems \cite{Zhang2012}.

\begin{figure}[h]
\includegraphics[scale=0.5]{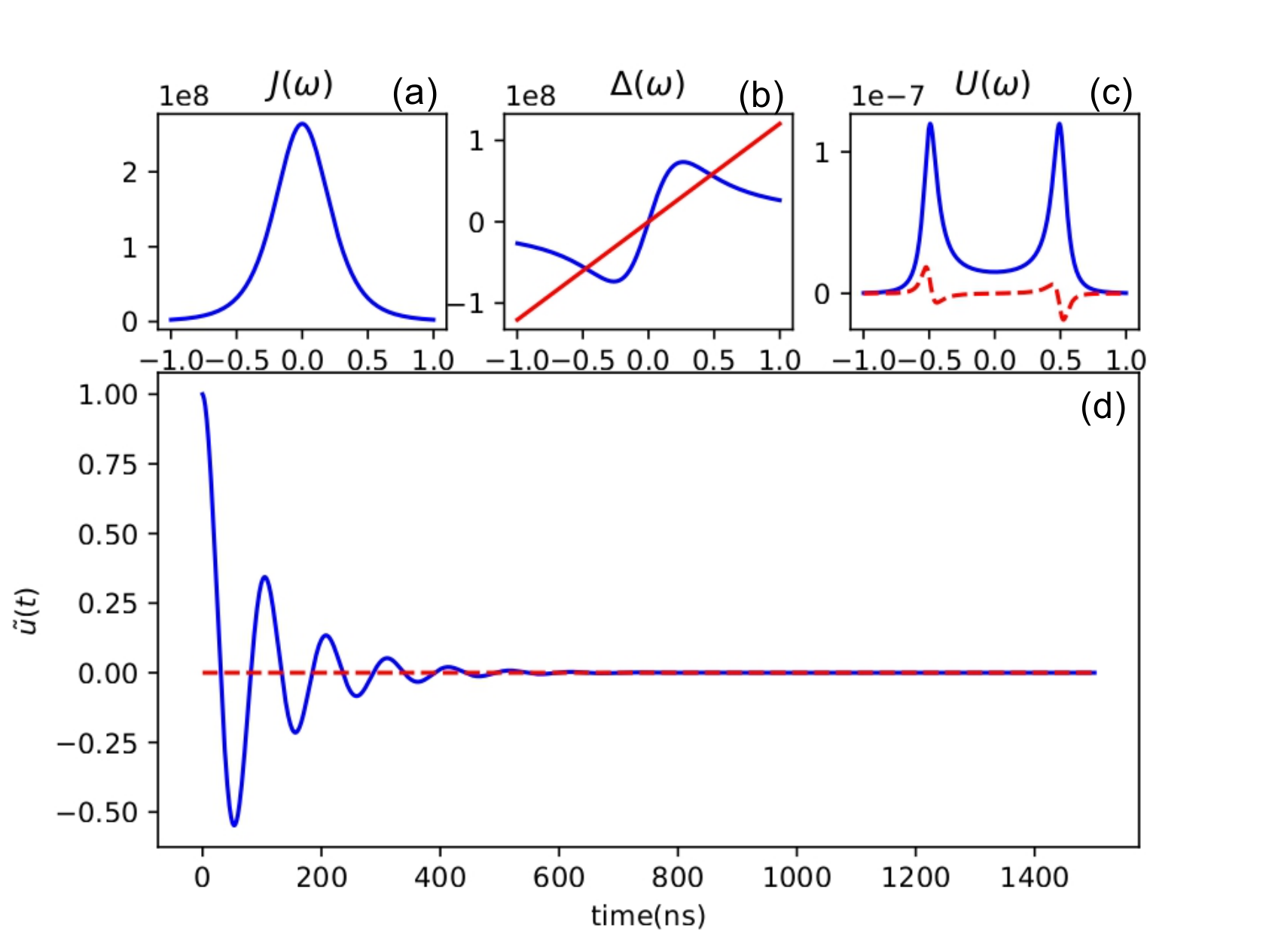}
\caption{(a) The q-Gaussian spectral density $J(\w)$ with respect to the frequency $\w$ in the unit of $\Omega_R$
(after a central frequency shift $\w-\w_s \rightarrow \w$, with  $\w_s=\w_c$), which is also the imaginary part 
of the self-energy. (b) The blue curve is the  real part of the self-energy, the cross points with the red line is the solution
of $\Delta(\w){=}\w$ at which $ J(\w)\neq 0$ so that there is no localized bound states.
(c) The blue (real) and red (dot)  curves are the real and imaginary parts of $U(\omega)$, respectively.
(d) The  inverse Laplace transformation of $U(\omega)$, i.e., the Green function $ \tilde{u}(t,t_0)$. Here we plot 
the real  and imaginary parts [the blue (real) and the red (dot ) curves] of  $\tilde{u}(t,t_0) \equiv u(t,t_0)  e^{i\omega_c(t-t_0)}$ with $t_0{=}0$.} 
\label{U_qG}
\end{figure}

 Because the cavity decay rate $\kappa \neq 0$, strictly speaking, no localized mode can exist in this hybrid quantum system. 
 Thus, the solution to $u(t,t_0)$ is simply given by
\begin{equation}
u(t{-}t_0)=\frac{2}{\pi} \!\! \int_{-\infty}^{\infty}\! \!\! d\w\frac{J_{s}(\w)\e^{-i\w (t-t_0)}}
    {4\big[\w-\w_c-\Delta(\w)+i\kappa\big]^{2} +J^2_{s}(\w)}.  \label{usolution}
\end{equation}
Figure \ref{U_qG} shows the detailed solution of the Green function $u(t,t_0)$ with the q-Gaussian spectral density 
Eq.~(\ref{sesd1}) for the spin ensemble. It contains all the decoherence information of the cavity system strongly coupled
with the spin ensemble, as shown in Fig.~\ref{U_qG}(d). These results are independent of the specific initial state of the cavity.

 Experimentally, the photon number is measured by detecting the transmission intensity of the cavity field. Our theory gives the solution 
 to the cavity photon number, i.e., Eq.~(\ref{cavityi}) 
which is divided into four terms. The first term represents the dissipation of the initial photons in the cavity. 
The second term is the photons injected into the cavity through the driving field. The third term shows the photon quantum fluctuations  
induced by the spin ensemble and the environment.  The last term indicates the interference between 
the initial cavity photons and injected photons.
In the experiment \cite{Putz14}, the cavity system is initially set in the vacuum. Hence, the first and the last terms are 
equal to zero in Eq.(\ref{cavityi}) so that it is simply reduced to
\begin{align}
\label{N}
\langle n(t) \rangle & =|y(t,t_0)|^{2} + v(t,t),
\end{align}
where $y(t,t_0)$ and $v(t,t)$ are given by Eq.~(\ref{yvt}) which is determined by the solution of the Green function  Eq.~(\ref{usolution}).
\begin{figure}[h]
\includegraphics[scale=0.5]{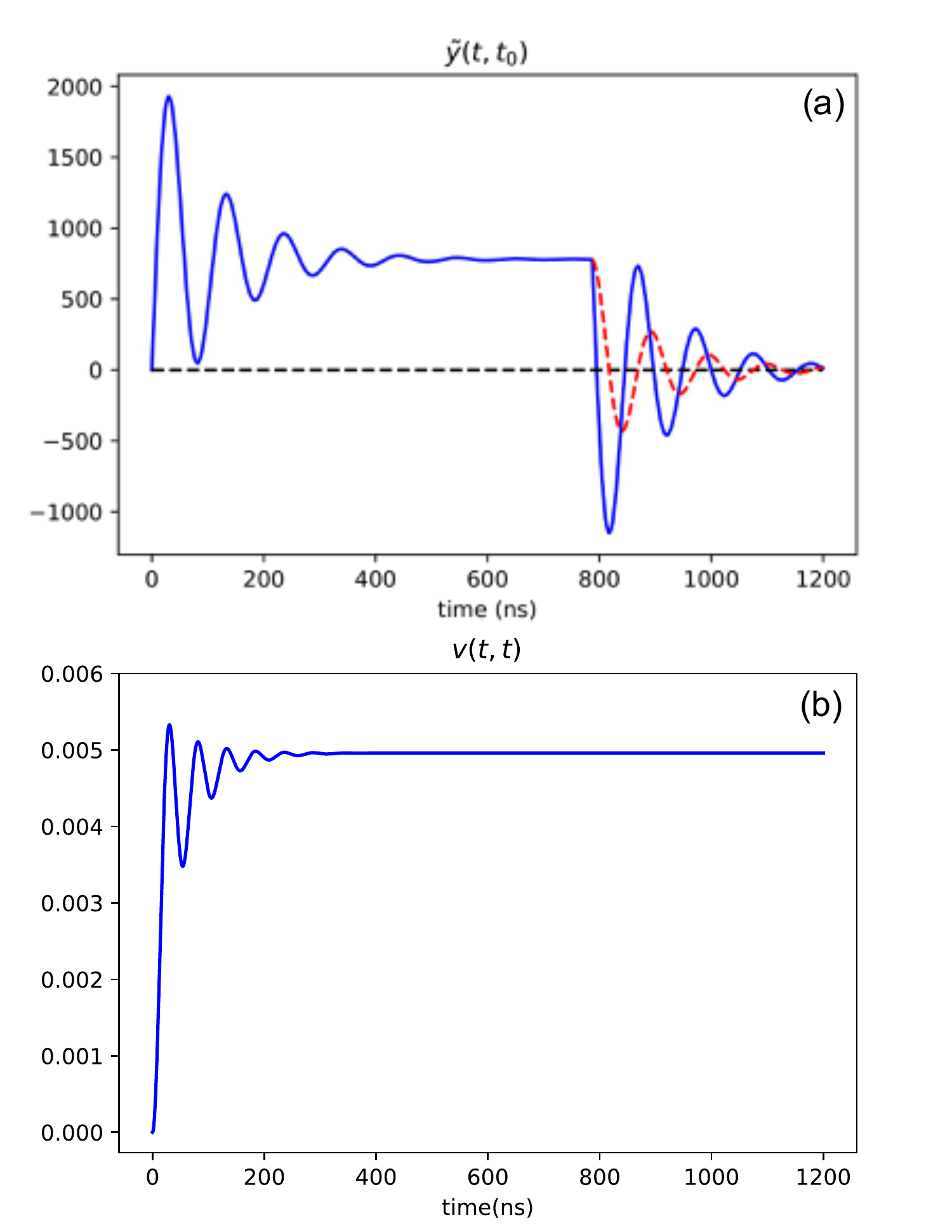}
\caption{(a) The blue line is the evolution of the driving-field-induced cavity field $\tilde{y}(t,t_0){=}y(t,t_0)\e^{i\w_{c}(t{-}t_0)}$, 
in which the dampings are caused by the spin ensemble for the q-Gaussian spectral density, and a rectangular driving field is 
applied with the resonant frequency $\w_p=\w_c=\w_s$. The red-dashed line is given by Eq.~(\ref{ey_off})
for the time duration $ t \geq t_{\text{off}}$. (b) The photon fluctuation $v(t,t)$ of Eq.~(\ref{vtt}) 
induced by spin ensemble at temperature 25 mK, which is negligible in comparison with the
driving-field-induced cavity field $|y(t,t_0)|^2$.}
\label{v_qG}
\end{figure}

The solutions of $y(t,t_0)$ and $v(t,t)$ are plotted in Fig.~\ref{v_qG} in which $\tilde{y}(t,t_0){=}y(t,t_0)\e^{i\w_{c}(t{-}t_0)}$. 
It shows that the driving-field-induced cavity field 
undergoes two damping processes. The first damping takes place when photons are injected into the cavity 
through a rectangular  driving field.
The cavity photons are raised up to the maximum very quickly. Meantime they are dissipated into the spin ensemble 
which is described  Eq.~(\ref{edf}) through the Green
function $u(t,t_0)$, the corresponding oscillation decay is similarly shown in Fig.~\ref{U_qG}(d).  The decay lasts a certain time $t_s$ 
($t_s < 600$ ns  for the parameters given in \cite{Putz14}), and the cavity photons gradually reach a saturation, 
i.e., $y(t_s,t_0) \simeq y(t_{\text{off}}) = \langle a^\t (t_{\text{off}})\rangle $ after $t_s \simeq 600$ ns.   
Also note that the oscillating decay is a  manifestation of the non-Markovian cavity decoherence induced by the inhomogeneous 
broadening of the spin ensemble spectrum, as we will see more later.  

\begin{figure}[h]
\includegraphics[scale=0.45]{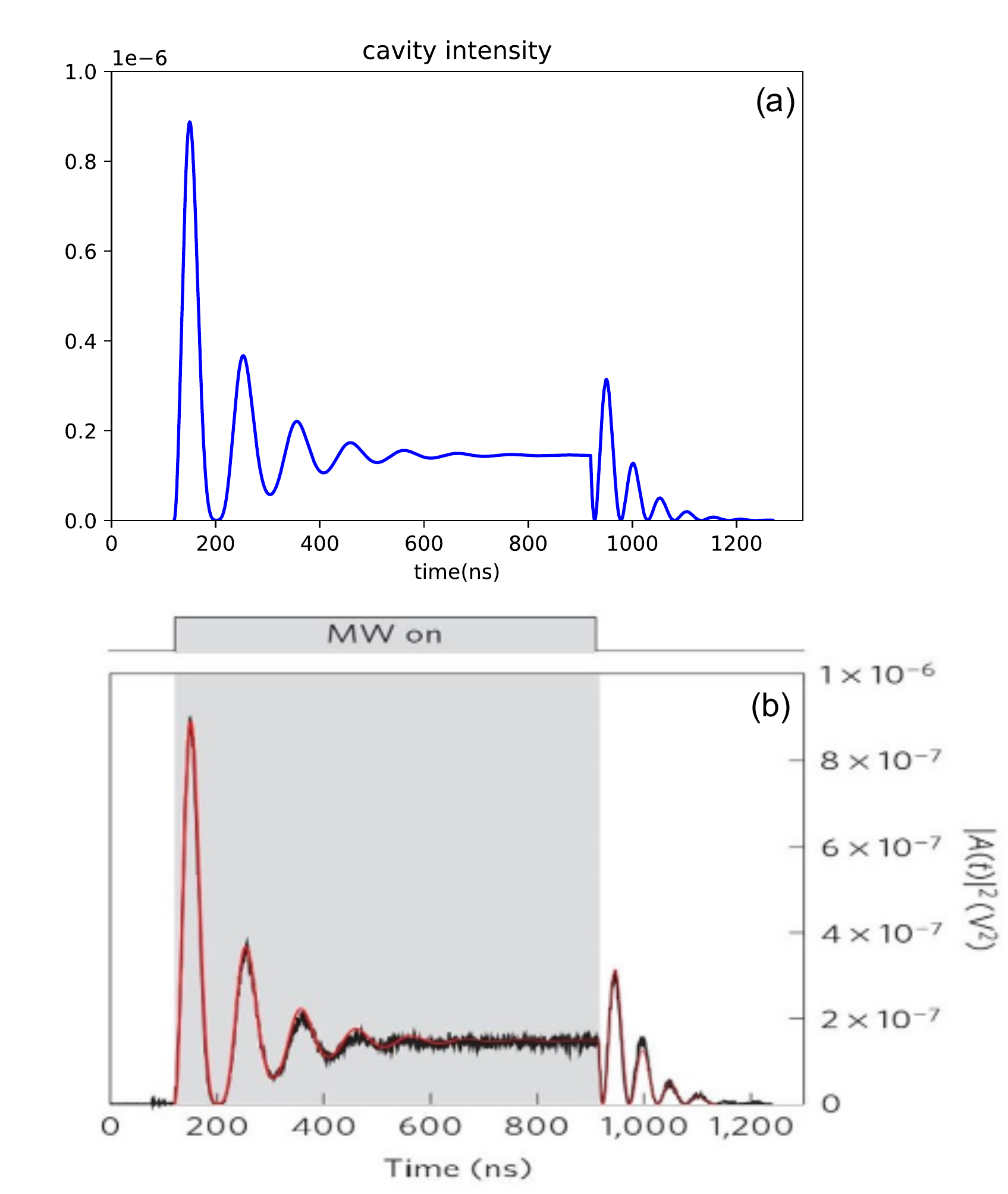}
\caption{(a) The cavity intensity under a rectangular driving field calculated from our theory. 
(b) The cavity intensity measured (black curve) in experiment and the semiclassical description (red curve) given in \cite{Putz14}. }
\label{N_qG}
\end{figure}

After the driving field is turned off at the time $t_{\text{off}}$, the cavity field undergoes the second damping, as shown 
by the blue solid line in Fig.~\ref{v_qG}(a). 
This result is directly obtained from the solution of $y(t_{\text{off}}+\tau,t_0)$ which can be reduced to
\begin{align}
\tilde{y}(t_{\text{off}}{+}\tau,t_0) & = \tilde{y}(t_{\text{off}}){-}\tilde{y}(\tau,t_0),  \label{y_off}
\end{align}
which shows that the second damping process is off-phase with the first damping process. As a result, the time scale of the second damping 
in the observation of the cavity intensity is only a half of the  decay time of the first damping, as shown in Fig.~\ref{N_qG}.  On the other hand, 
the second damping process contains two parts, one is the re-started damping of the cavity field $y(t_{\text{off}})$ after the driving field is 
turned off, which is given by
\begin{align}
\tilde{y}_1(t_{\text{off}}{+}\tau,t_0) 
& =  \tilde{y}(t_{\text{off}}) \tilde{u}(\tau, t_0) .
\label{ey_off}
\end{align}
The result is shown by the red-dashed line in Fig.~\ref{v_qG}(a). The difference between the blue-solid line and the red-dashed line 
in Fig.~\ref{v_qG}(a) is the field retrieved from the spin ensemble. This part is  the field retrieved back from
the spin ensemble, as a demonstration of  quantum memory in this hybrid systems \cite{Putz14}.
\begin{figure}[h]
\includegraphics[scale=0.55]{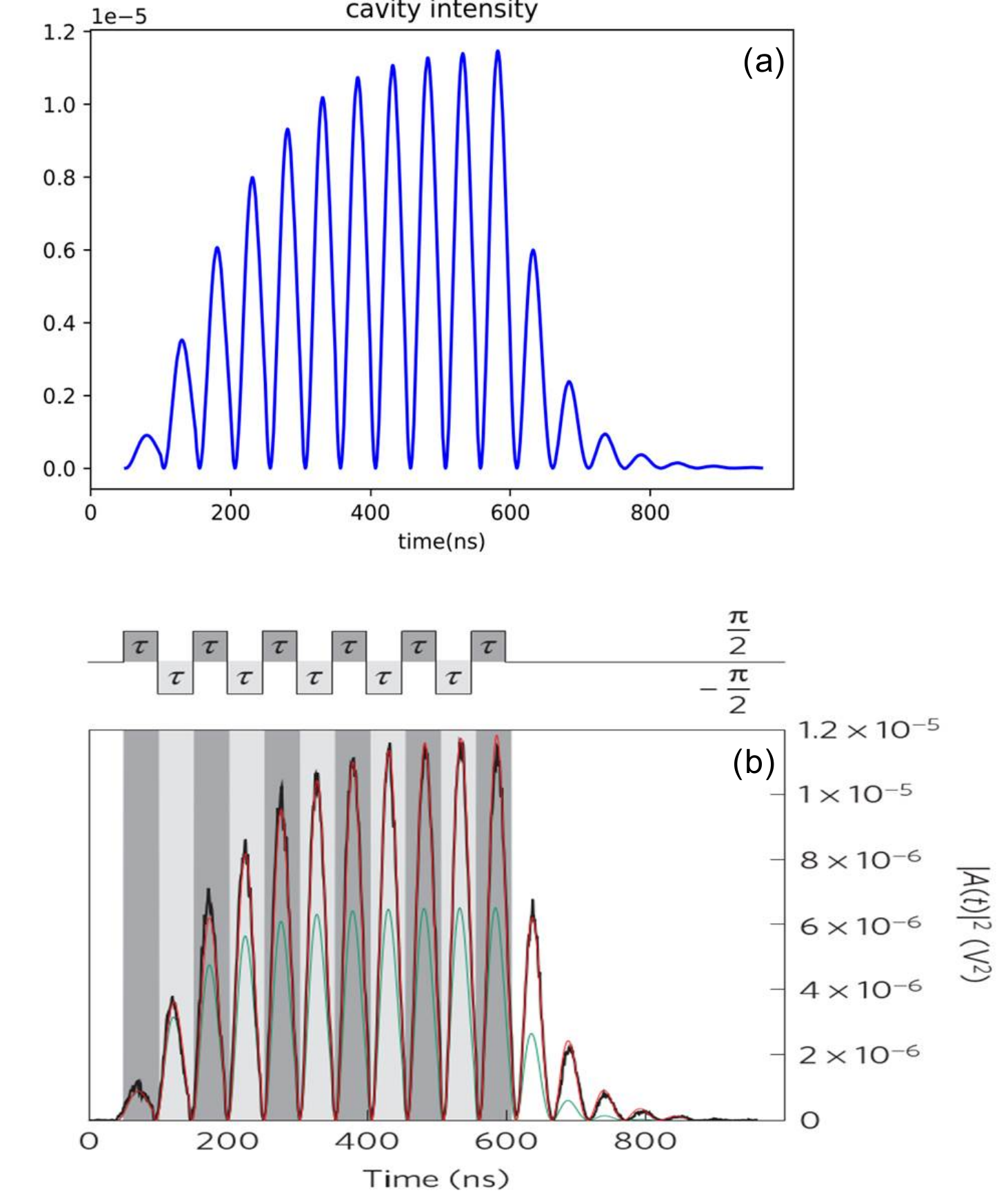}
\caption{(a) The cavity intensity under a phase-shifting driving field calculated from our theory. (b) 
The cavity intensity measured (black curve) in experiment and the semiclassical description (red curve) presented  in \cite{Putz14}.}
\label{N_pi}
\end{figure}

With the above picture how the cavity field induced by the driving field is decohered (stored) into the spin ensemble
and how it is retrieved back from the spin ensemble after the driving field is turned off, we present in Figs.~\ref{N_qG}(a) 
and \ref{N_pi}(a)  the cavity intensity of Eq.~(\ref{N}) with a rectangular driving field 
and a $\pi$ phase-shifting driving field, respectively, observed in the experiment \cite{Putz14}.
Figures~\ref{N_qG}(a) and \ref{N_pi}(a) show that our theoretical predictions are in good 
agreement with the experimental data presented in Figs.~\ref{N_qG}(b) and Fig.~\ref{N_pi}(b) of Ref.~\cite{Putz14}. In fact, the authors of
Ref.~\cite{Putz14} used a semiclassical approach (the Volterra integral equation) to describe the cavity photon dynamics, and their results 
 (see the red curves in Figs.~\ref{N_qG}(b)  and Fig.~\ref{N_pi}(b)) also fitted the experimental data very well.  This is because, 
 when the injected photons through the driving field  is very large ($\approx 10^6$) and the temperature is set to be very low ($\approx 25 mK \sim 
 \frac{1}{5} \hbar\w_{c}$) in the  experiment, the photon fluctuations given by $v(t,t)$ is indeed very very small, namely, 
 $v(t,t) \ll |y(t,t_0)|^2$, as shown in Fig.~\ref{v_qG} [comparing the magnitude difference between Fig.~\ref{v_qG}(a) and Fig.~\ref{v_qG}(b)].
This gives the reason why the semiclassical approach used  in \cite{Putz14} can fit very well with their experimental data.

\begin{figure}[t]
\includegraphics[scale=0.5]{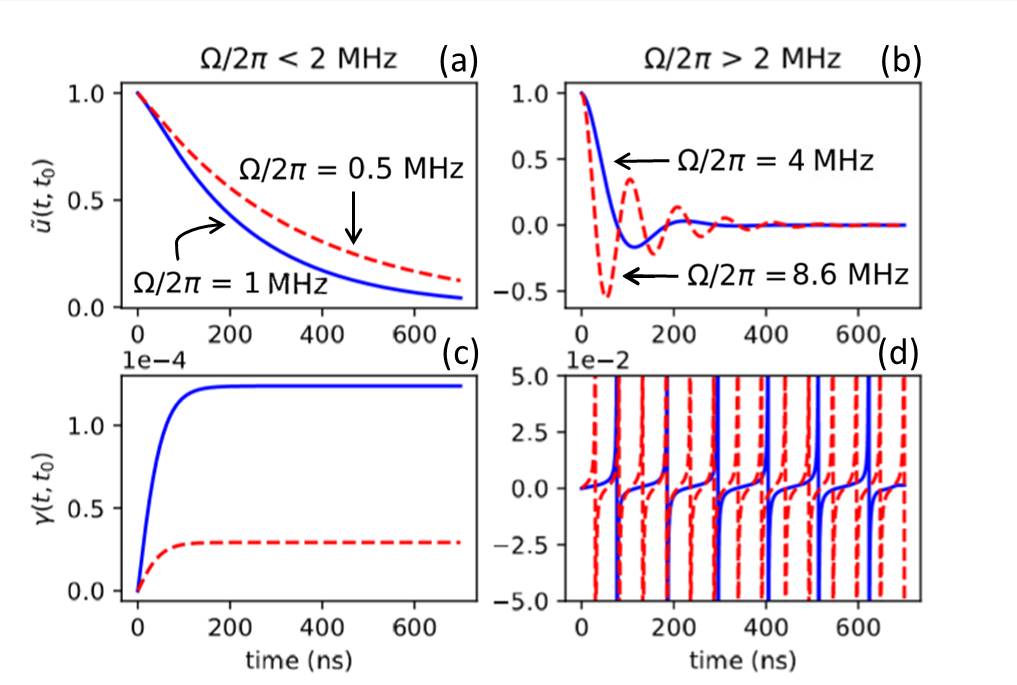}
\caption{ Decoherence dynamics of the cavity photons dissipating into the spin ensemble for different 
coupling strength $\Omega$ between the cavity and the spin ensemble, the same other parameters are
taken as given in the experiment \cite{Putz14}. (a)-(b) The time evolution of the Green 
function $u(t,t_0)$ for the coupling strength $\Omega/2\pi=0.5, 1.0$ MHz and $4, 8.6$ MHz, respectively;
(c)-(d) The corresponding dissipation coefficient $\gamma(t,t_0)$ determined by Eq.~(\ref{dissipation}).}
\label{Gamma_s}
\end{figure}

On the other hand, the oscillation decays in Figs.~\ref{N_qG} and \ref{N_pi} manifest the non-Markovian decoherence dynamics of the cavity field 
induced by the spin ensemble, here the cavity leakage is very small ($\kappa = 2\pi \times 0.4 $ MHz
$\ll \Omega = 2\pi \times 8.6$ MHz) in the experiment setup \cite{Putz14}. 
These oscillation decays can be characterized by the dissipation coefficient $\gamma(t,t_0)$ in the exact master equation 
Eq.~(\ref{M_eq}) and can be  explicitly computed through Eq.~(\ref{dissic}) which is again 
determined by the Green function $u(t,t_0)$ of Eq.~(\ref{usolution}),
\begin{align}
\gamma(t,t_0)= - {\rm Re}[\dot{u}(t,t_0)/u(t,t_0)] .   \label{dissipation}
\end{align}
Some results are plotted in Fig.~\ref{Gamma_s}.  Taking the same parameters used in experiment \cite{Putz14}, we find that 
when the coupling strength $\Omega/2\pi < 2$ MHz, 
the Green function $u(t,t_0)$ shows an exponential decay, see Fig.~\ref{Gamma_s}(a), which is the Markovian decoherence process.
The corresponding decay rate is given by the asymptotic value of $\gamma(t,t_0)$, see Fig.~\ref{Gamma_s}(c). The asymptotic value 
is $\gamma(t \rightarrow \infty,t_0) \simeq 2 \kappa + J_s(\omega_s)$, as expected in the experiment observation in \cite{Putz14}. 

\ \\

For the strong coupling regime (here $\Omega/2\pi > 2$ MHz), the decoherence dynamics is very different. 
The Green function $u(t,t_0)$ is no longer an exponential decay,
it involves an oscillation decay for the cavity field as shown in Fig.~\ref{Gamma_s}(b). Such oscillations lead to the stronger oscillation 
of the dissipation coefficient $\gamma(t,t_0)$ in time, as shown in  Fig.~\ref{Gamma_s}(d). As it is well-known, such an oscillation
shown in the dissipation coefficient between the positive and negative value is the strong effect of non-Markovian decoherence 
dynamics. In particular, the sudden change in the dissipation coefficient from a huge positive value to a huge negative value 
corresponds to the rapid forward and backward flows of photons (and information) between the cavity and the spin ensemble in a 
very short time. As the coupling becomes stronger and stronger, such rapid forward and backward flows of photons (or information) 
between the cavity and the spin ensemble get faster and faster. It is this unusual non-Markovian dynamics suppresses the cavity 
photon decoherence, as observed in this hybrid quantum system \cite{Putz14}. Also, such non-Markovian phenomena usually occurs 
in the symmetric spectral density with respect to the system energy state. This is the underlying mechanism how 
the decoherence induced by the inhomogeneous broadening is suppressed  in the strong-coupling regime.  
This rapid forward and backward flows of photons 
between the cavity and the spin ensemble play a similar role of dynamical decoupling by a rapid, time-dependent control modulation 
to suppress decoherence \cite{DD1,DD2} even though the mechanisms of decoherence suppression are completely different. 
The former comes from the symmetric spectral density where the spectral density is the decoherence source, 
while the latter is generated from the external control pulses.  Therefore, symmetricalized spectral densities may provide 
an alternative way to suppress decoherence for structured environments. 

\section{Decoherence suppression by hole-burning spectra}
\begin{figure}[h]
\centering
\includegraphics[scale=0.5]{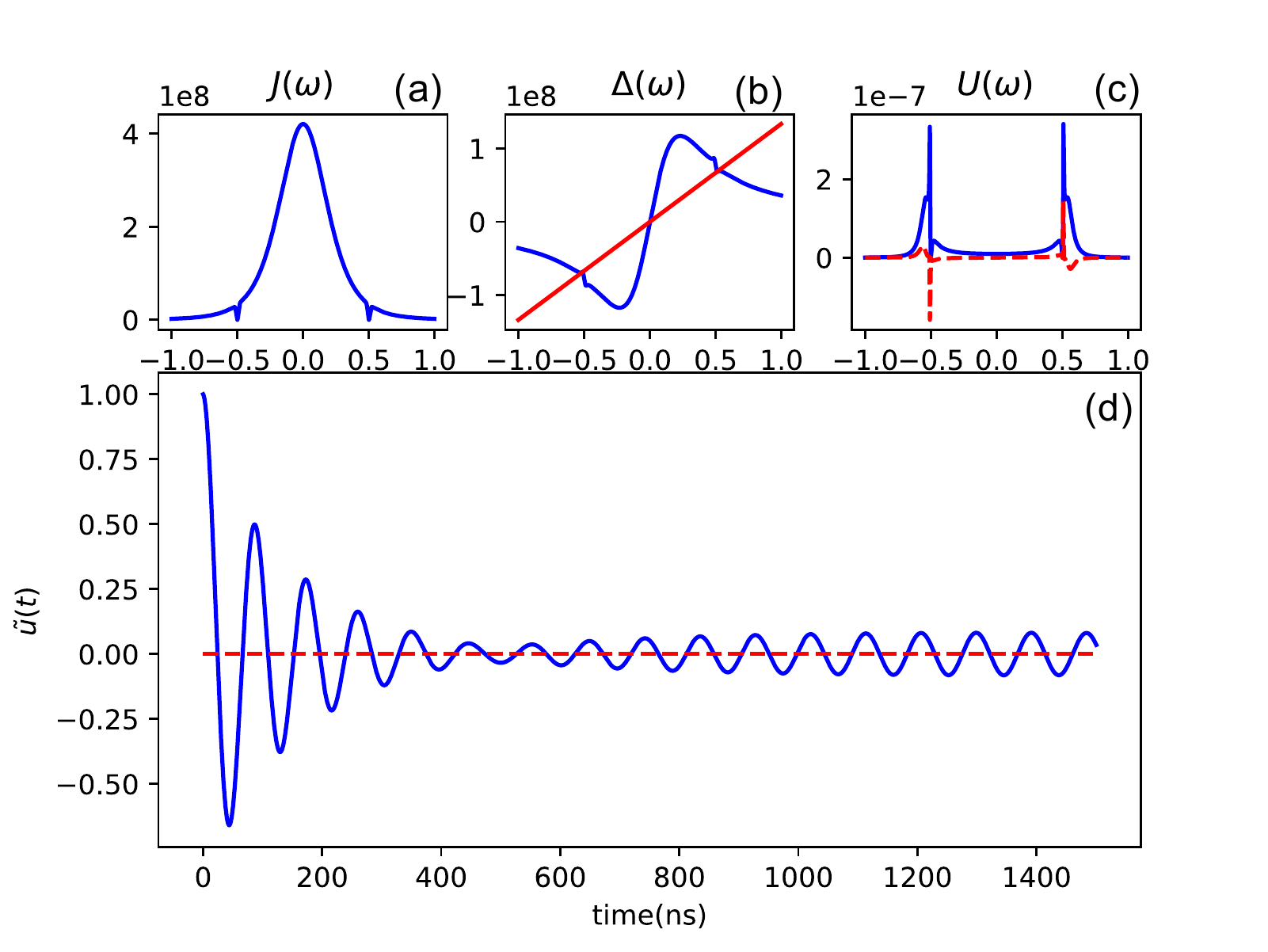}
\caption{(a) The spectral density after the spectral hole burning.
(b) The blue curve is the  real part of the corresponding self-energy, the cross points with the red line are the solutions
of $\Delta(\w){=}\w $, corresponds to the localized mode frequencies $\w_b=\w_{s}{\pm}\frac{\Omega_R}{2}$.
(c) The real part (blue real line) and the imaginary part (red dot line) of the the spectrum of $u(t,t_0)$. 
(d) The real part (blue real line) and the imaginary part (red dot line) of $\tilde{u}(t,t_0)$ with $t_0{=}0$.}
\label{U_HB}
\end{figure}

In another experiment of Putz {\it et al.} \cite{Putz17}, the spectral density of the spin ensemble is modified by the spectral hole burning technique.
The spectral density after hole-burning frequencies at $\w_{s}{\pm}\frac{\Omega_R}{2}$ is plotted 
in Fig.~\ref{J_qG}(a) [also see Fig.~\ref{U_HB}(a)]. In Fig.~\ref{U_HB}(b), the intersection points of the self-energy $\Delta(\omega)$
with the red line are located at frequencies $\w_{s}{\pm}\frac{\Omega_R}{2}$, which could result in two localized modes \cite{Zhang2012}
by the condition $\omega-\omega_s-\Delta(\omega)=0$ if the cavity leakage $\kappa$ can be ignored. 
It should be pointed out that the localize modes are the bound states of the cavity incorporating all possible modes of the spin ensemble. 
These localized modes are actually different from the dark states in quantum optics.
Correspondingly, in Fig.~\ref{U_HB}(c), the two sharp peaks at $\w_{s}{\pm}\frac{\Omega_R}{2}$  
show the positions of the two localized modes, respectively. As a result,  the Green function $u(t,t_0)$ 
keeps oscillation in time, coming from these two localized modes, as shown in Fig.~\ref{U_HB}(d). 
In other words, the cavity field will not reach a steady state with the spin ensemble, as a dissipationless 
process induced by the localized modes due to the hole burning spectral density. 
This could change significantly the decoherence dynamics of the cavity field. 
Note that in principle, the peaks resulting from localized modes in Fig.~\ref{U_HB}(c) should be $\delta$-functions. 
However, due to the existence of a small cavity leakage ($\kappa \neq 0$), the peaks have a small finite width 
which results the cavity field in a slow decay in a very long time which is not shown in the Fig.~\ref{U_HB}(d).

\begin{widetext}

\begin{figure}
\includegraphics[scale=0.65]{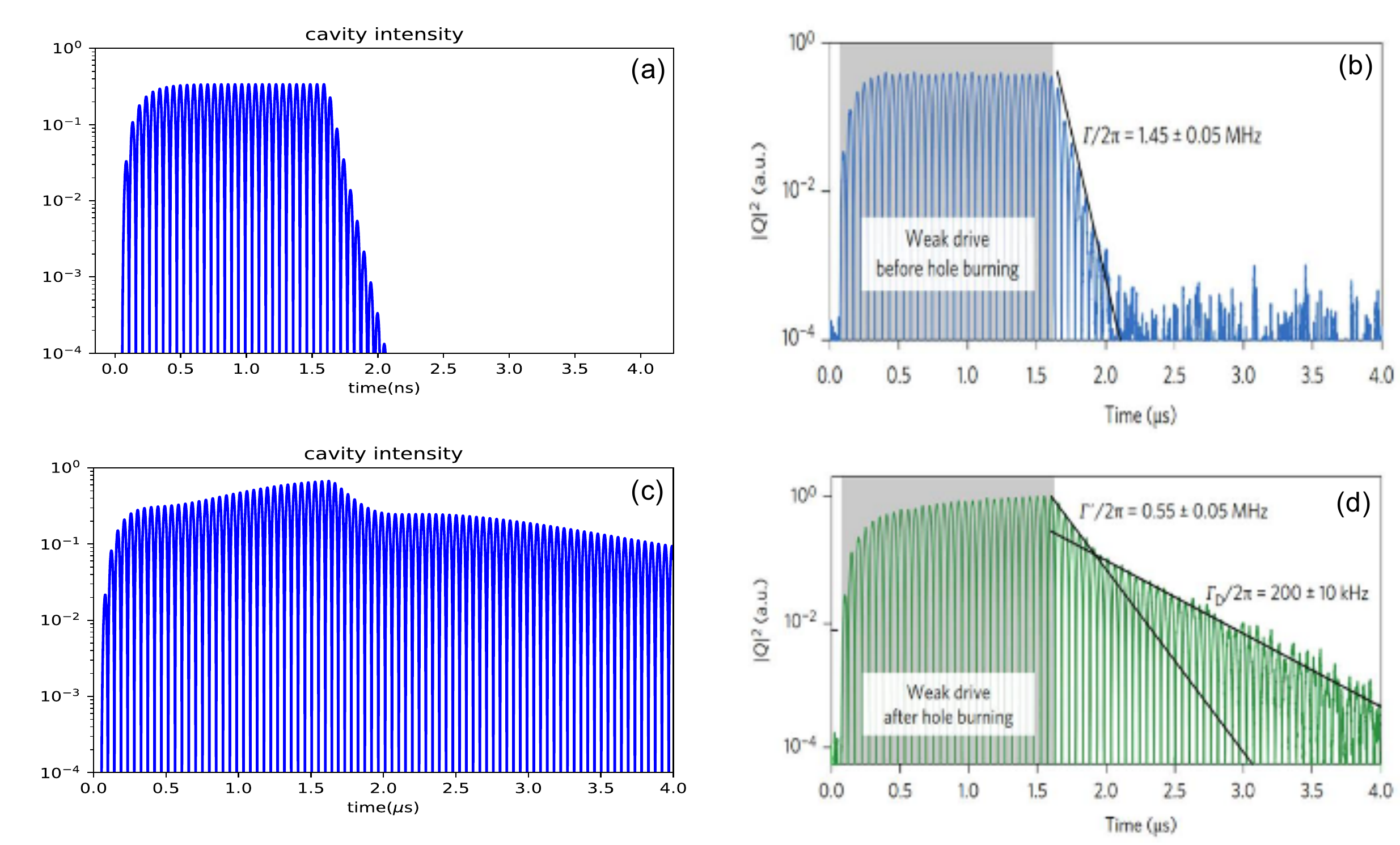}
\caption{Cavity intensity for the q-Gaussian spectral density without (a-b) and with (c-d) the spectral hole burning, under a sinusoidal driving field, in which (a) and (c) the theoretical 
calculation from our theory; (b) and (d) are the experimental measurement by Putz {\it et al.} \cite{Putz17}. }
\label{N_sin.qG}
\end{figure}

\end{widetext}

In the experiment \cite{Putz17}, the cavity transmission intensity is measured by applying a sinusoidal modulated pulse. 
Figure \ref{N_sin.qG} shows a comparison of the cavity intensity with and without the spectral hole burning 
on the spin ensemble spectrum. Both the theoretical calculations and the experimental results show that, 
without the spectral hole burning, the cavity intensity has a linear decay in time after the driving field 
is turned off, see  Fig.~\ref{N_sin.qG}(a) and (b).  While, with the spectral hole burning to the spin ensemble
spectrum, the damping of the cavity intensity slows down significantly after the driving field 
is turned off, see  Fig.~\ref{N_sin.qG}(c) and \ref{N_sin.qG}(d). In other words, the coherence time is
substantially improved. This is an effect of the dissipationless localized modes induced by the spectral hole burning 
of the spin ensemble. 

\begin{figure}
\includegraphics[scale=0.5]{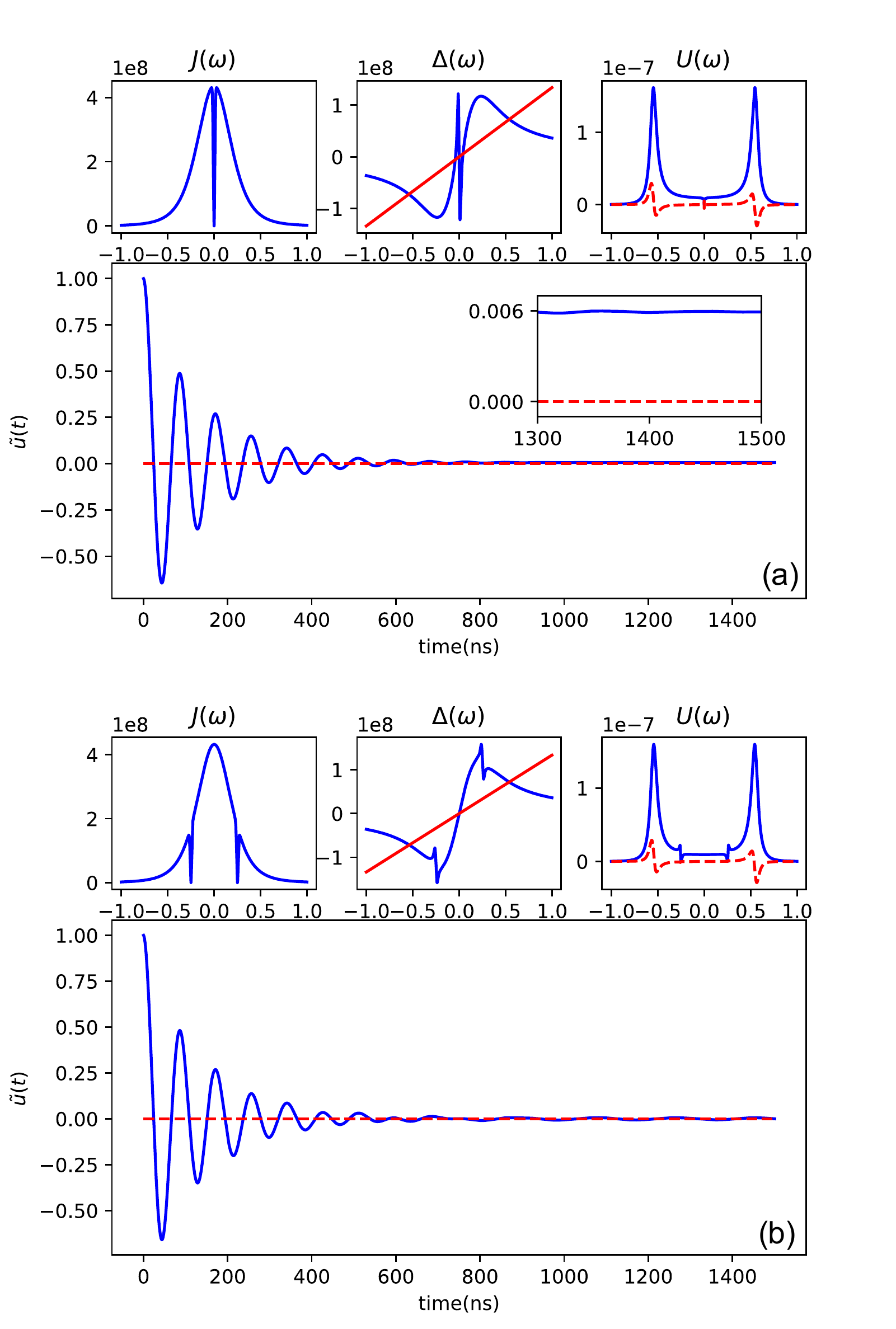}
\caption{ The same plot as Fig.~\ref{U_HB} with different burning frequencies. (a) burning frequency at $\omega= \omega_s$.
(b) A pair burning frequencies at $\omega= \omega_s{\pm}\frac{\Omega_R}{4}$.}
\label{U_HB00}
\end{figure}

Note that the theoretical and the experimental results shown in Fig.~\ref{N_sin.qG}(c)-(d) do not fit each other very well.  
This is because the decoherence dynamics is very sensitive to the position and the shapes of the burning holes. 
In addition to the spectral hole burning at frequencies given in \cite{Putz17}, we also consider burning frequencies at different position to see how  
 the decoherence dynamics of the cavity field sensitively depends on the  shape of the spectral density.
 In Fig.~\ref{U_HB00}(a), the spectrum is burnt at $\omega_s{=}\omega_c$.  Theoretically, there may  exist a localized mode at frequency $\omega_s$ if the
  cavity decay $\kappa$ can be negligible so that the decoherence can be suppressed. 
  However, because the slope of $\Delta(\w)$ at $\omega_s$  [see Fig.~\ref{U_HB00}(a)] is so steep such that the amplitude of this localized mode 
  is very small. In other words, the burning frequency almost has no effect on the cavity decoherence dynamics, as shown in the inset of Fig.~\ref{U_HB00}(a). If the spectrum is burnt at 
  $\omega_s{\pm}\frac{\Omega_R}{4}$ as shown in Fig.~\ref{U_HB00}(b). 
 Qualitatively, the decoherence dynamics is no much difference in comparing with the case without the hole burning, see in Fig.~\ref{U_qG}(d), although the decay is slow down a little bit. 
  This shows that the burning frequencies must locate at the position of the intersection points with the red line in Fig.~\ref{U_HB00}(b) so that
  the dissipationless localized modes can exist. 
  
  In summary, theoretically the suppression of cavity decoherence or the improvement of the transmission cavity field 
  through the spectral hole-burning technique are resulted from the dissipationless localized modes
  between the system and the spin ensemble with gaped spectral density. In particular, these two localized modes are 
 collective  bound states (with many photons and spins) localized in energy domain \cite{Zhang2012}, which is quite 
 different from usual dark states interpretation given  in \cite{Putz17}. 
 These states can be used to store (trap) information but are not easy to retrieve the information back.  Therefore, it may not be useful for quantum 
  memories but it can be used to serve as qubit states for quantum information processing, because these states can be made as decoherence-free states
  with specific design of spectral hole burning to the spin or atomic ensembles, even for heat ensembles.
 
\section{Characterizing Quantum memory with two-time correlation functions}

 As one sees, dynamical properties of the cavity system are fully determined by the Green function $u(t,\tau)$, from which
 one can find how the driving-field-induced cavity field  $y(t,t_0)$ changes inside the cavity through 
 Eq.~(\ref{yvt}). It contains all the information on cavity decoherence induced by the spin ensemble through the
 spectral density $J_s(\omega)$. In this section, we show how the memory effect between the cavity and the spin ensemble can be
explored specifically. To investigate the memory effect, we shall explore the two-time correlation functions, which measure
 the correlation of a past event with its future and therefore provide a physically observable  definition of memory \cite{Ali2015}. Meantime, the two-time correlation function 
is now experimentally measurable in the real-time domain. In quantum optics, the familiar two-time correlation functions are the first-order
 and the second-order coherence functions, defined by
 \begin{align}
 g^{(1)} (t,t+\tau) &= \frac{\langle a^\dag(t) a(t+\tau) \rangle}{\sqrt{\langle a^\dag(t) a(t) \rangle \langle a^\dag(t+\tau) a(t+\tau)\rangle}} , \\
g^{(2)}(t,t+\tau) &=\frac{\braket{a^{\t}(t)a^{\t}(t+\tau)a(t+\tau)a(t)}}              {{\braket{a^{\t}(t)a(t)}}\braket{a^\t(t+\tau)a(t+\tau)}},
 \end{align}
 which can be calculated in terms of the functions $u(t,t_0)$, $v(\tau,t)$ and $y(t,t_0)$ in out theory. 
 Explicitly,
\begin{align}
\label{FCF}
\braket{a^{\t}(t)a(t+\tau)} = & u^{*}(t,t_0)u(t+\tau,t_0)\braket{a^{\t}(t_0)a(t_0)} \notag \\
& +y^{*}(t,t_0)y(t+\tau,t_0) + v(t+\tau,t) \notag \\
                        & +u^{*}(t,t_0)y(t+\tau,t_0)\braket{a^{\t}(t_0)}  \notag \\
                        &    +u(t+\tau,t_0)y^{*}(t,t_0)\braket{a(t_0)}.
\end{align}
The second-order correlation function is more complicated although it can still be expressed 
in terms of the basic Green  functions $u(t,t_0)$, $v(\tau, t)$ and $y(t,t_0)$.  

Here we only present the result $\braket{a^{\t}(t)a(t+\tau)} $ to demonstrate the memory effect in this 
hybrid quantum system. Because the cavity is initially empty, the above solution is reduced to
\begin{align}
\label{FCF}
\braket{a^{\t}(t)a(t+\tau)} =  y^{*}(t,t_0)y(t+\tau,t_0)  + v(t+\tau,t) . 
\end{align}
It shows that the coherence function is divided into two parts,  the first part (the first term) 
is related to the coherence of the driving-field-induced cavity field at two different 
times, and the second term is associated with spin-ensemble-induced photon correlation. The cavity
intensity presented in the previous two sections is the special case with the delay time $\tau=0$.
In Fig.~\ref{N_qG}, it shows that the driving-field-induced cavity photons dissipate into the spin ensemble  
in a finite time scale, then the cavity field reaches a saturation (steady state). After turning off the driving 
field, the remained photons in the cavity continuously dissipate into the spin ensemble, mixed with some 
photons retrieved back from the spin ensemble. The question
to ask is how these photons coherent with the photons previously injected into the cavity through the driving field.

\begin{figure}
\includegraphics[scale=0.52]{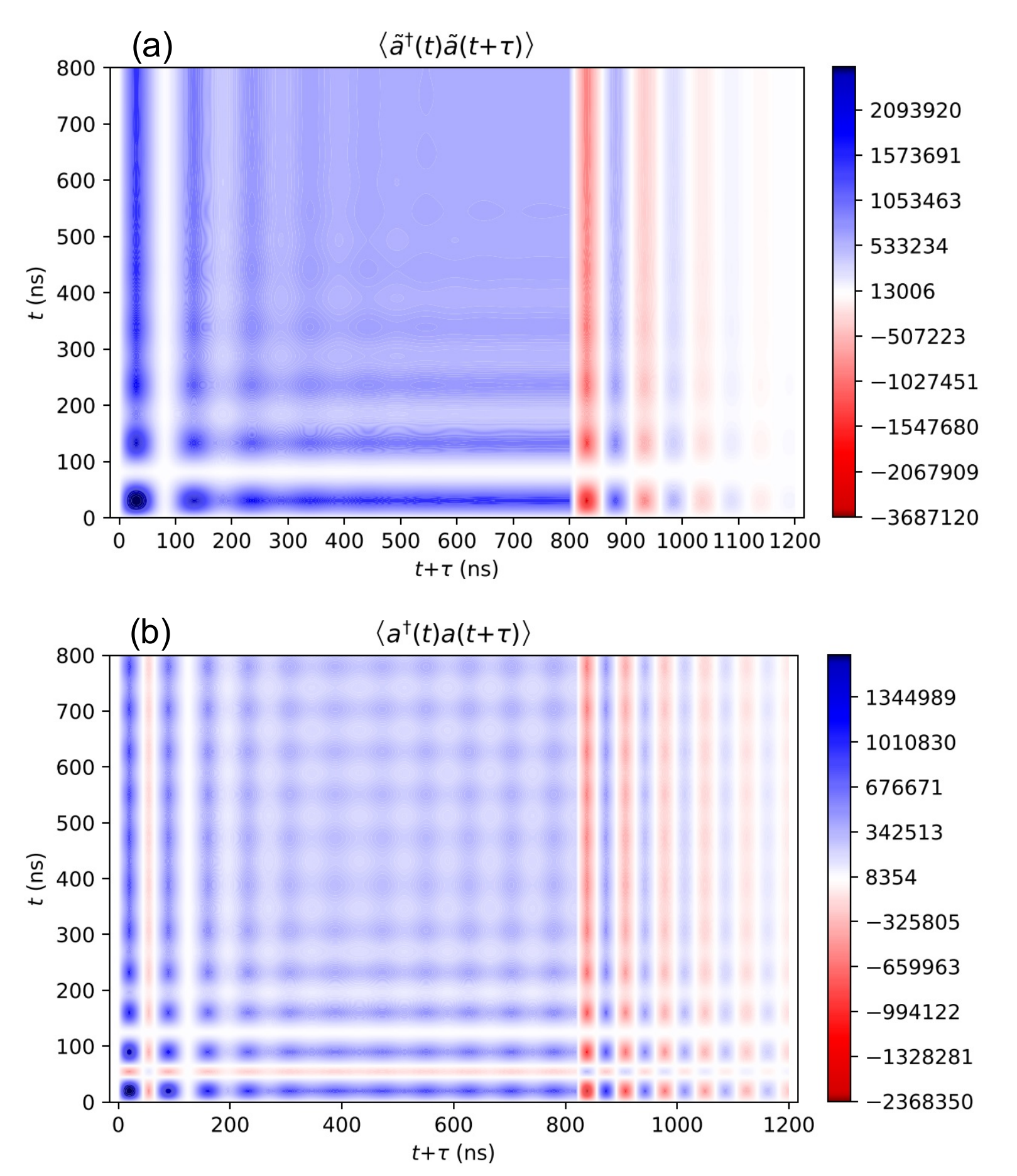}
\caption{First-order correlation (coherence) function $\braket{\tilde{a}^{\t}(t)\tilde{a}(t+\tau)}$ under the rectangular driving field (a) in the q-Gaussian spectral density 
and (b) with the spectral hole burning.}
\label{FCF_qG}
\end{figure}

 Figure \ref{FCF_qG}(a) shows the first-order coherence function $\braket{a^{\t}(t)a(t+\tau)}$ of the cavity field 
 for a rectangular driving field with the same parameters used in Fig.~\ref{N_qG}. The result shows that the coherence between 
 time $t$ and $t{+}\tau$ decays gradually before the driving field is turned off. However, once the driving field is turned off, the value of 
 the first-order coherence drops to be a negative value. This negative correlation corresponds to an opposite $\pi$ phase coherence between 
 the cavity fields at time $t$ and $t{+}\tau$, respectively. This is clearly shown by Eq.~(\ref{y_off}),
 and is manifested significantly in Fig.~\ref{FCF_qG}(a). Moreover, we have pointed out that this $\pi$-phase shift 
of the cavity field at a later time turns out a reduction of the damping oscillation period time by a half, which is also clearly seen in Fig.~\ref{FCF_qG}(a).
On the other hand, Fig.~\ref{FCF_qG}(b) shows how the coherence function is changed by the spectral hole burning.
The coherence behavior with the  spectral hole burning is obviously lasted much longer, which indicates that the spectral 
hole burning can effectively suppress the decoherence and  maintain the cavity field coherence for longer time.

\begin{figure}
\includegraphics[scale=0.52]{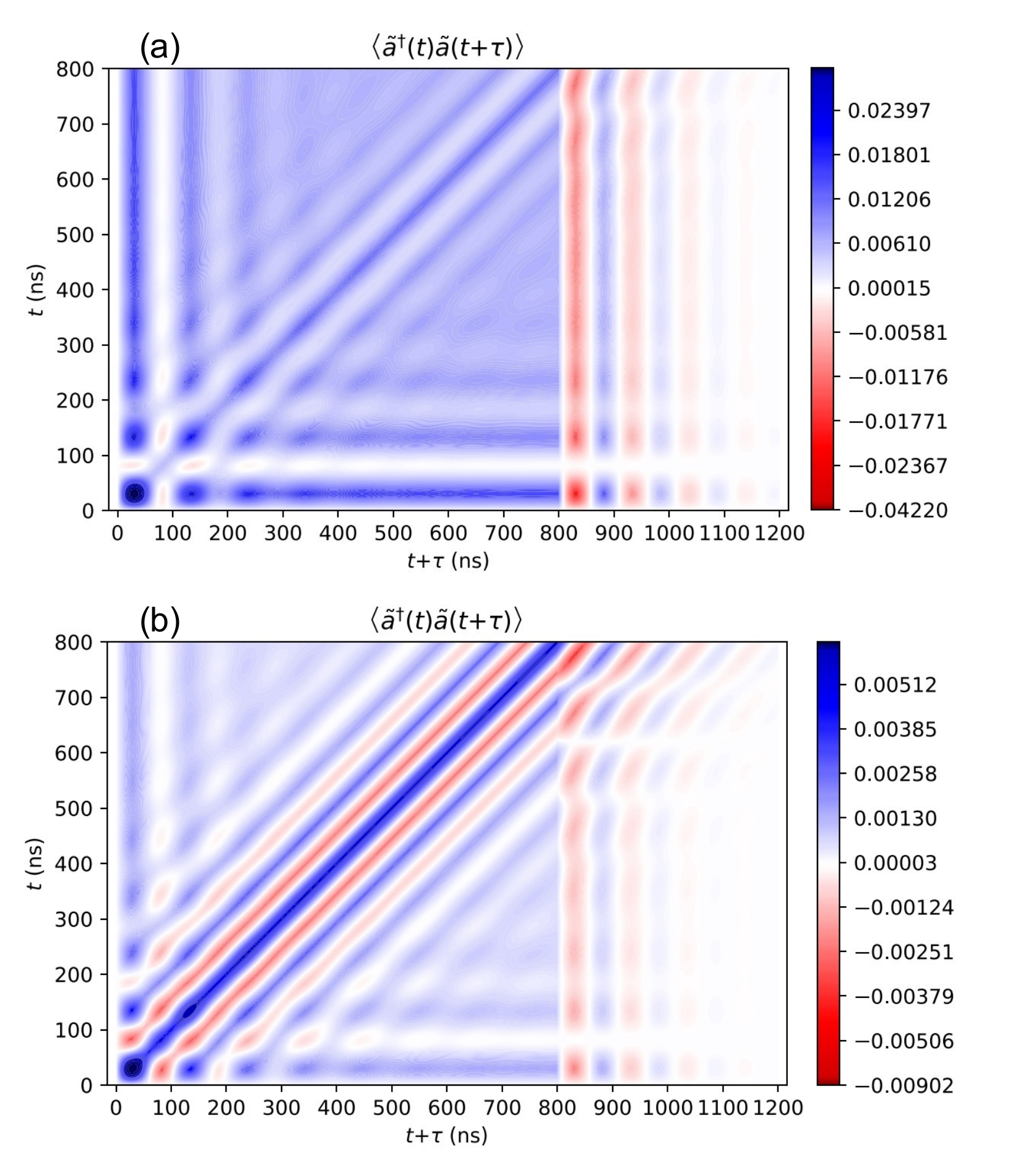}
\caption{The same plot as Fig.~\ref{FCF_qG} with a deduced driving field amplitude by an order of $10^{-4}$.}
\label{FCF_8}
\end{figure}

However, the coherence functions shown in Fig.~\ref{FCF_qG} are acturaly dominated by the classical coherences or classical correlations
because the cavity field contains a large number of photons (the order of $\sim 10^6$) which is in classical limit. 
To make the quantum coherence manifest, one should use a very weak driving field that maybe only contain one 
photon or less. The results are presented in Fig.~\ref{FCF_8}. 
As one can see, after reduced down the driving field pulse by a order of $10^{-4}$, 
the two-time correlations are significantly changed. This change is mainly because the quantum noise correlation given
by $v(t,t+\tau)$ becomes no longer negligible.   The diagonal fringes in Fig.~\ref{FCF_8} are the manifestation of this 
quantum noise correlation. 

\begin{figure}
\includegraphics[scale=0.52]{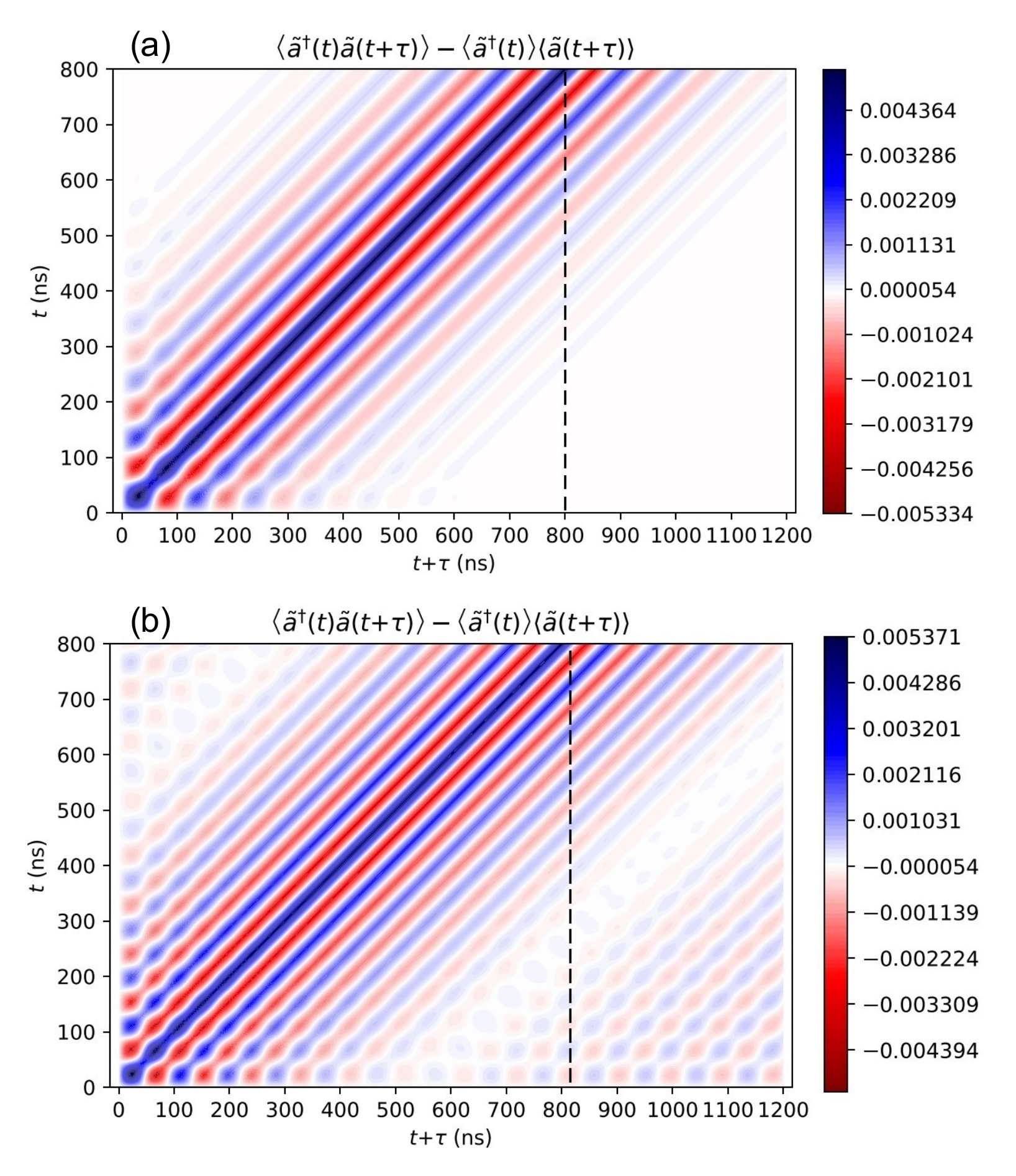}
\caption{(a) First-order quantum correlation function in the q-Gaussian spectrum.
(b) First-order quantum correlation function with the spectral hole burning.}
\label{FQCF_qG}
\end{figure}

To see more clear the quantum correlations, we plot the noise correlation alone by subtracting the classical part
\begin{align}
\label{FQCF_LT}
 \braket{a^{\t}(t)a(t+\tau)}-\braket{a^{\t}(t)}\braket{a(t+\tau)}  
= v(t+\tau,t)
\end{align}
The result is presented in Fig.~\ref{FQCF_qG}. Actually the quantum noise correlation
itself is independent from the driving field, it manifests the quantum memory effect between the cavity system and 
the spin ensemble, and is sensitive to the spectral structure of the spin ensemble.  Figure \ref{FQCF_qG}(a) and (b) 
shows a clear difference of the quantum correlations between  the q-Gaussian spectrum and 
the corresponding spectral hole burning. As one can see from
Fig.~\ref{FQCF_qG}(a) that without spectral hole burning, there is no obvious quantum correlation for the injected cavity field with cavity field 
after the driving field turned off. But with the spectral hole burning, the spin-ensemble-induced decoherence is 
suppressed and quantum memory can be maintained, which is displayed 
by the fringes in the right-bottom corner in Fig.~\ref{FQCF_qG}(b). It shows a long time correlation or a long time quantum 
memory effect in this hybrid quantum systems, which should be interest to measure in further  experiments.

\section{Discussions and Outlook}

In conclusion, we provide a master equation approach to study the decoherence dynamics of the hybrid quantum systems 
consisting of a microwave cavity coupled with an inhomogeneous broadening spin ensemble. The spin ensemble is made by 
negatively charged nitrogen-vacancy (NV) defects in diamond. Under the condition that less spins ($\sim 10^6$) are excited for  
a large spin ($\sim 10^{12}$) ensemble, the spin ensemble can be treated by a bosonic ensemble 
under the Holstein-Primakoff approximation, and from which we can derive the exact master equation for the cavity field.
Our exact master equation describes the transient dynamics 
of the cavity system under the control of the external driving field  in both the classical and the quantum regimes.  
The study of non-Markovian decoherence dynamics of the cavity coupled strongly to the spin ensemble  
is reduced to solve the Green functions $u(t,t_0)$, $v(\tau,t)$ and $y(t,t_0)$, i.e.~Eqs.~(\ref{yvt}) and (\ref{usolution}). 
These Green functions can be analytically solved and the results depict explicitly the detailed dissipation, quantum 
correlations and quantum memory effects in such hybrid quantum systems. 

We apply the theory to the recent experiments by Putz et al. \cite{Putz14,Putz17}
and interpret the corresponding experimental observation. Although the experimental observations have also been 
described with the semiclassical mean-field approach in terms of the Volterra integral equation in \cite{Putz14,Putz17}, 
or can be described by other approaches, such as the approaches presented in Refs.~\cite{Kurucz11,Diniz11,Sandner11}, 
our results lead to a clearer physical picture of the decoherence. In particular, we find that the suppression of the
decoherence induced by the inhomogeneous broadening of the spin ensemble in the strong-coupling regime is
due to the rapid forward and backward photon flows between the cavity and the spin ensemble manifested in terms
of the time-dependent dissipation coefficient in the master equation. The rapid forward and backward photon flows 
are also a typical non-Markovian decoherence effect. Such a rapid forward and backward photon flows become faster  
with the stronger coupling strength, manifested by the rapid oscillation of dissipation coefficient 
from the huge positive value to the huge negative value, see Fig.~\ref{Gamma_s}(d). It plays a similar role of dynamical 
decoupling by a rapid, time-dependent control modulation for decoherence suppression \cite{DD1,DD2} but it comes from 
the symmetric spectral density with respect to the cavity frequency rather than the external 
control pulses. This is the mechanism how the decoherence is suppressed in the strong-coupling regime in cavity QED. 
Moreover, the further suppression of decoherence through spectral hole burning \cite{Putz17} is due to the fact that the hole burning spectral 
density generates localized states which is dissipationless, as we pointed out early \cite{Zhang2012}.

We further explore the quantum memory in this hybrid quantum system through the two-time correlation functions. 
From a measurement perspective, the correlation between past events and their future is a measurable 
memory feature. We calculate the first-order coherence correlation function, and show how the cavity field correlated 
each other before and after the driving field turned off. In particular, we show how the quantum correlation or quantum 
memory can be manifested in the correlations function in the quantum regime with a low photonic intensity, which
should be not difficult to be measured in experiments. Furthermore, based on our master equation, we can 
solve explicitly the reduce density matrix of the cavity field in this system, from which one can examine how the quantum 
states are stored in the spin ensemble and retrieved in a later time through quantum state tomography. 
This will remain in our further research.

\begin{acknowledgements}
This work is supported by Ministry of Science and Technology of Taiwan under Contract No. NSC-108-2112- M-006-009-MY3.
\end{acknowledgements}

\end{document}